\documentclass[a4paper]{article}
\usepackage[T1]{fontenc}
\usepackage{lmodern}

\usepackage{geometry}
\usepackage{authblk}
\usepackage{setspace}
\usepackage[pdftex]{graphicx}
\graphicspath{{./figures/}}
\usepackage{subcaption}
\usepackage{amsmath, amssymb, mathrsfs}
\usepackage{float}
\usepackage{balance}
\usepackage{enumitem, array}
\usepackage{multirow, booktabs}
\usepackage{tabularx}
\usepackage{comment}
\usepackage{adjustbox}
\usepackage{booktabs}
\usepackage{caption}
\usepackage{xcolor}
\usepackage[table]{xcolor}
\usepackage{booktabs}
\usepackage{float}
\usepackage{mhchem}
\usepackage{siunitx}
\usepackage[colorlinks=true, citecolor=blue, linkcolor=blue, urlcolor=blue]{hyperref}

\usepackage[
    style=numeric-comp, 
    sorting=none,       
    backend=biber,      
    defernumbers=true,
    maxcitenames=2,
    giveninits=true
]{biblatex}
\usepackage{lineno}

\usepackage[labelfont=bf]{caption}
\usepackage{longtable} 
\usepackage{booktabs}  
\usepackage{array}

\title{Microfluidic Study of Evaporation-Driven Crystallization of Saline and Ammonia Brines under Hydrogen Flow}

\author[1*]{Karol M. D\k{a}browski}
\author[2]{Mohammad Nooraiepour}
\author[3]{Mohammad Masoudi}
\affil[1]{\small{Faculty of Drilling, Oil and Gas, AGH University of Krakow, al. Mickiewicza 30, 30-059, Krakow, Poland}}
\affil[2]{\small{Environmental Geosciences, Department of Geosciences, University of Oslo, P.O. Box 1047 Blindern, 0316 Oslo, Norway}}
\affil[3]{\small{SINTEF Industry, Applied Geoscience Department, 7465 Trondheim, Norway}}
\affil[*]{\small{Corresponding author: karol.dabrowski@agh.edu.pl}}

\begin{document}
\maketitle

\begin{abstract}
Underground storage of hydrogen and ammonia in geological formations is essential for renewable energy integration, but salt precipitation during gas injection may threaten storage performance. While extensively studied for CO\textsubscript{2} systems, precipitation mechanisms in hydrogen-brine and ammonia-brine systems remain poorly understood. This study presents a comprehensive microfluidic investigation of salt crystallization during hydrogen injection into saline and ammonia-containing brines using high-pressure microfluidics. We conducted 81 high-pressure experiments systematically varying brine composition (1-5 mol/kg NaCl), chemical additives (surfactants, alcohols, ammonia), and hydrogen flow rates (200-1300 mL/min). Quantitative image analysis reveals that hydrogen-induced precipitation differs fundamentally from CO\textsubscript{2} systems. Hydrogen drives physical precipitation via evaporation and capillary trapping, producing discrete, localized deposits. In contrast, CO\textsubscript{2}-ammonia systems generate extensive reactive precipitation of ammonium bicarbonate with interconnected crystal networks. Interfacial tension (IFT) controls both residual brine distribution and final crystal coverage: high-IFT fluids form large, interconnected brine pools promoting extensive crystallization, while low-IFT fluids create isolated pools reducing crystal coverage by 50\%. Alcohol and surfactant additives suppress precipitation by enhancing brine mobility, whereas ammonia paradoxically increases crystal fractions despite lower IFT.  Higher flow rates accelerate crystallization across all compositions, enabling operational mitigation strategies.  and demonstrate that gas-specific, rather than CO\textsubscript{2}-analog, risk assessments are essential for underground hydrogen storage design. The effectiveness of chemical additives offers promising pathways for near-wellbore protection in underground hydrogen storage operations.\\

\end{abstract}

\section{Introduction}
\label{Introduction}

The global energy transition to net-zero carbon emissions requires widespread adoption of renewable energy sources and significant emissions reductions to meet ambitious climate targets, such as the EU's Fit for 55 package, which aims to reduce carbon dioxide emissions by 55\% by 2030.\cite{EU2021Fit55,bistline2021roadmaps,nooraiepour2025norwegian, schlacke2022implementing}.  However, the intermittent and seasonal nature of renewable energy production from solar and wind sources creates significant challenges for grid stability and energy security \cite{bandi2017spectrum,shi2020fluctuation,heide2010seasonal}. This variability necessitates large-scale, long-term energy storage solutions capable of balancing energy supply and demand across temporal scales ranging from hours to seasons \cite{kiviluoma2012short,haydt2011relevance,pascual2021energy}.

Among the various energy storage technologies, chemical energy carriers, particularly hydrogen and ammonia, have emerged as leading candidates for large-scale (\SI{\sim100}{MWh} \cite{caglayan2020technical}), long-term (weeks to months) energy storage \cite{kharel2018hydrogen, zhang2016survey}. Hydrogen offers high energy density and chemical stability, while ammonia provides additional advantages, including higher volumetric energy density (\SI{12.7}{\mega\joule\per\liter}) and simpler storage requirements (readily liquefied at \SI{\sim10}{\bar} or \SI{-33}{\degreeCelsius}) \cite{arsad2022hydrogen, chang2021emerging}. Ammonia presents an excellent volumetric hydrogen density, with approximately \SI{108}{\kilogram\per\cubic\meter} of H\textsubscript{2} embedded in liquid ammonia at \SI{20}{\degreeCelsius} and \SI{8.6}{\bar}, compared to advanced hydrogen storage systems like metal hydrides, which store H\textsubscript{2} up to \SI{25}{\kilogram\per\cubic\meter}.

These energy carriers can be produced during periods of renewable energy surplus and utilized for power generation when renewable production is reduced \cite{turner2008renewable, acar2014comparative}. Additionally, ammonia has the unique capability to serve as a CO\textsubscript{2}-free energy storage medium, unlike other hydrogen carriers such as methanol or formic acid, which release carbon oxides during their use. Both hydrogen and ammonia can serve dual purposes in industrial applications, particularly in fertilizer production and other chemical processes \cite{ghavam2021sustainable}.

Surface hydrogen storage facilities are limited and costly, making subsurface hydrogen storage in geological formations a more viable alternative due to the substantial capacity, safety, and economic feasibility. Underground storage is essential for accommodating the massive volumes required for national and regional energy systems \cite{zivar2021underground,masoudi2024lined,tarkowski2019underground,park2024underground}.

Currently, hydrogen is predominantly stored in salt caverns, which offer excellent properties including low permeability, high sealing capability, inert chemical behavior with respect to hydrogen, and favorable mechanical properties for repeated injection-withdrawal cycles \cite{ozarslan2012large,tarkowski2018salt}. Salt caverns are widely investigated for energy storage, allowing for the secure storage of fluids over extended periods. However, the availability of suitable salt formations is geographically constrained \cite{slizowski2017potential}.

Saline aquifers and depleted gas fields offer compelling alternatives to salt caverns for energy storage, owing to their greater geographical distribution and substantial storage capacity \cite{tarkowski2017perspectives,raad2022hydrogen,nooraiepour2025geologicalco2storageassessment,sainz2017assessment}. However, these formations present additional technical challenges due to the low density of hydrogen compared to natural gas, requiring significantly larger pore volumes to store equivalent amounts of energy.

Ammonia has a higher energy density than hydrogen and benefits from established infrastructure, including pipelines, shipping, and handling systems already existing globally. Despite these advantages, ammonia storage faces significant safety constraints due to its toxicity, limiting storage options primarily to caverns. Even in these controlled environments, structural leakage in inter-layers remains a concern \cite{ghaedi2025characterization}.
A particularly complex scenario arises in co-storage applications where hydrogen and ammonia may interact within the same geological formation. In salt caverns where ammonia is stored as a hydrogen carrier, subsequent hydrogen injection could displace stored ammonia into surrounding rock interlayers. This displacement could trigger localized ammonia-brine interactions, potentially leading to accelerated evaporation and complex precipitation processes that remain poorly understood.

When gases are injected into saline formations, they displace resident brine, leading to evaporation of residual brine under reservoir conditions \cite{nooraiepour2018effect,nooraiepour2025three}. This evaporation causes salt precipitation, a phenomenon well-documented in CO\textsubscript{2} storage and natural gas operations \cite{miri2016salt,nooraiepour2018salt}. Salt precipitation in porous media can significantly impair storage operations by forming pore-blocking salt crystals that reduce the porosity and permeability, as well as the injectivity, of near-wellbore regions and the overall reservoir formation \cite{nooraiepour2025three,masoudi2021pore,masoudi2024mineral}. In severe cases, precipitated salts can weaken the rock matrix, compromising the mechanical integrity of the reservoir \cite{nooraiepour2025potential,aminzadeh2025ultrasonic}.

The mechanisms of salt precipitation in CO\textsubscript{2}-brine systems have been thoroughly studied through laboratory and numerical experiments, as well as several field observations \cite{talman2020salt, miri2015new}. However, the behavior in hydrogen-brine and ammonia-brine systems remains poorly understood \cite{gholami2023hydrogen, ghaedi2025evaporation}, with a significant lack of pore-scale mechanistic insights. This knowledge gap is critical because hydrogen and ammonia differ significantly from CO\textsubscript{2} in solubility, reactivity, and transport properties, which may strongly influence brine displacement, salt precipitation patterns, and consequences for flow percolation pathways.

Microfluidic experiments have been instrumental in elucidating mineral dissolution and precipitation reactions under precisely controlled conditions, offering high-resolution, pore-scale insights into these processes \cite{nooraiepour2018effect, fazeli2019microfluidic,lei2025advancing,dkabrowski2025surface,zhang2025pore,nooraiepour2021probabilistic,roman2025microfluidics,jiang2025controls}. Microfluidic chips enable precise control of experimental conditions, allowing for real-time monitoring of dynamic processes. The small volumes inherent to these systems accelerate evaporation and salt crystallization \cite{zhang2024brine}, making microfluidics an optimal platform for studying complex interactions in hydrogen-ammonia-brine systems under controlled, observable conditions.

Despite the growing importance of hydrogen and ammonia for energy storage \cite{chang2021emerging}, experimental investigations of hydrogen-brine and ammonia-brine salt precipitation systems remain scarce. The fundamental mechanisms controlling salt precipitation, their dependence on fluid properties, and their potential impacts on reservoir performance in hydrogen-ammonia storage scenarios are not yet well understood. This represents a critical knowledge gap that must be addressed to evaluate the long-term feasibility and safety of underground hydrogen and ammonia storage operations.

In this work, we aim to extend microfluidic experimental capabilities to explore scenarios where hydrogen is injected into chips pre-saturated with brine and ammonia-brine mixtures. This experimental design enables us to mimic both hydrogen-brine displacement in saline aquifers and the complex situation in salt caverns where hydrogen displaces stored ammonia, potentially driving ammonia-brine evaporation and accelerating salt precipitation.
By systematically varying gas flow rates, brine compositions, and ammonia concentrations, we quantify how displacement dynamics, interfacial tension, and evaporation kinetics control precipitation pathways in these systems. These experiments provide the first direct visualization of salt precipitation under hydrogen-ammonia subsurface storage conditions, yielding new insights critical for the safe and efficient implementation of underground energy storage technologies.

\section{Materials and Methods}

\subsection{Microfluidic chips and laboratory setup}
Crystallization experiments have been conducted using a microfluidic chip (Micronit Micro Technologies). The chip is composed of borosilicate glass with a thickness of 1800 $\mu m$ and 20 mm (length) $\times$10 mm (width) with total volume $18$ $\mu$l. The micromodel has laser-etched channels that resemble a realistic rock structure. The size of the pores varies from 100 $\mu$m to 600 $\mu$m, with a median value of 250 $\mu$m with average porosity $0.47\pm1$ and permeability $7.2\pm1.1$  $D$. The pore distribution exhibits a random pattern.
    
\begin{figure}
	\includegraphics[width=0.95\textwidth]{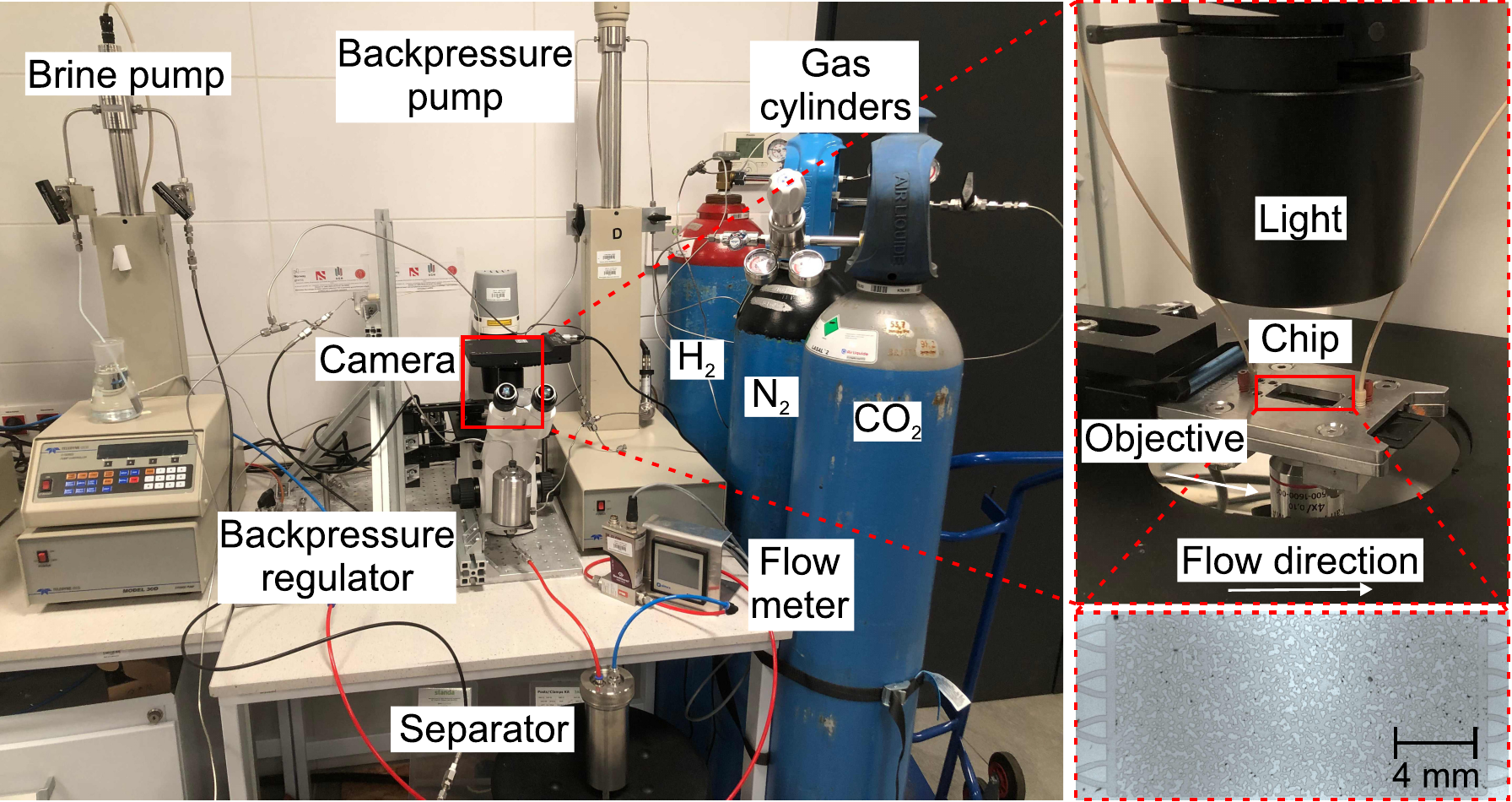}
	\caption{Photograph of high-pressure microfluidic setup utilized for crystallization process imaging. The setup consists of high-pressure syringe pumps used for brine injection and backpressure control, a high-speed microscopic camera used for image recording, a separator, and a flow meter. Chip is placed on the XYZ motorized stage. Flexible PEEK tubing allows chip moment. The first insert shows the tubing connection and chip holder. The second insert shows a microscopic image of the chip with a low-resolution objective. The setup is a modified version of the system used in \cite{dkabrowski2025surface,dkabrowski2025microfluidic}}
    \label{fig:ExperimentalSetup}
\end{figure}

The chip was mounted in a custom-made setup to facilitate in-situ observation of salt precipitation. Fig.~\ref{fig:ExperimentalSetup} shows a photograph of the setup used, while the insert shows an enlarged area of the chip holder, and the second insert shows the chip structure. The setup is a modified version of the setup used for salt precipitation studies under CO$_2$ flow \cite{dkabrowski2025surface}. The chip is mounted in a dedicated stainless steel holder (Micronit Micro Technologies) with a tubing connection. The frame is placed in the XYZ moving stage (Standa Ltd.) with a position resolution of 0.125 $\mu m$ for precise positioning during measurements. Flexible $1/16''$ PEEK tubing allows 20 mm movement in each direction. This allows us to imagine multiple positions within the chip network with high resolution. The upper working pressure of the system was 10 MPa. For imaging, a ZEISS Primo Vert microscope, equipped with a binocular phototube and a Plan-Achromat 4x/0.10 objective, is used. The images were recorded with a FASTEC IL5 high-speed camera with a 2560 x 2080 12-bit CMOS sensor, a 5 $\mu$ m low-noise pixel size and a frame rate of 20,000 fps. Connected to the microscope. Although the camera has a high frame rate, in these tests, due to the required high image resolution as well as the considerably low process time, the frame rate used was significantly lower. For a 4x objective, the field-of-view was a 2$\times$2 mm. The XYZ stage enabled the capture of a panoramic image of the chip, measuring 2$\times$20 mm in size. To saturate the chip, high-precision syringe pumps (Teledyne ISCO 30D) were used. The pressure was maintained with a backpressure regulator (Vinci Technologies BPR series 10000) connected to a syringe pump (Teledyne ISCO 100 DX). Flow was monitored with a Coriolis flow meter (IN-FLOW Bronkhorst). Temperature and pressure with differential pressure transducers (Keller PD-33 with temperature sensor). After the backpressure regulator, a custom-made two-phase separator was placed.  
  
For each measurement system, the chip was vacuumed and then saturated with a brine solution. The inlet tube was then disconnected and the dry gas tube was connected. This minimizes the volume of brine in the system and its contact with hydrogen (H$_2$). At this stage, the system at the input has no T-connectors or valves where residual brine could accumulate. The inlet gas is ensured to be dry, and the brine is not saturated with hydrogen. In the next step, the chip is pressurized using a gas cylinder to set the desired pressure. The backpressure regulator is controlled with a syringe pump that operates with an N$_2$. When the outlet pressure drops, the flow through the system begins. The pressure drop is regulated by the pump to maintain the set gas flow. In the current studies, the inlet pressure was set at 5 MPa. Note that the outlet pressure could vary in the event of a drop in chip permeability or clogging due to salt accumulation. The gas flow evaporates the brine on the chip. When the solubility limit is exceeded, a precipitate forms. Once the brine has completely evaporated, the measurement is finished. The experimental chip was then cleaned with deionized water and dried. Experiments were conducted for a set of brines and three hydrogen flow rates: 200, 400, and 1300 ml/min. For comparison, a single brine has been measured for a carbon dioxide (CO$_2$) flow rate of 400 ml/min. The crystallization processes of various brines during CO$_2$ injection have been investigated using a consistent experimental protocol, as detailed in \cite{dkabrowski2025microfluidic, dkabrowski2025surface}.

The interfacial tension (IFT) for gases and brines was measured using the pendant drop method \cite{berry2015measurement,higgs2022situ} with a high-pressure cell (Core Lab IFT Cell 1000). These measurements provide key parameters for understanding gas–liquid interactions inside the microfluidic chip. In particular, the IFT governs wetting and contact line dynamics, which affect how efficiently the injected gas displaces brine, how thin brine films remain before evaporation, and ultimately how and where crystallization initiates. Comparing H$_2$ and CO$_2$, differences in crystal growth and possible reaction of gas with a brine, which can rationalize variations in brine removal efficiency, saturation profiles, and crystallization onset observed in the flow-through experiments.

\subsection{Experimental fluid preparation}
We evaluated a total of nine fluids in the experiments, as listed in Table~\ref{tab:IFT_Flow}. Table~\ref{tab:IFT_Flow} shows the brine composition, their total dissolved sodium chloride (NaCl) concentrations, and IFT measured for hydrogen and carbon dioxide under 5 MPa. For clarity, the flow rates considered are shown. For each brine and flow rate, we conducted three experiments to ensure reproducibility, resulting in a total of 81 experiments. The first three brines (W1, W2.5, and W5) were prepared as aqueous stock solutions of NaCl dissolved in deionized water. Higher NaCl concentrations are expected to result in faster crystal formation and higher crystal fraction in the chip. It can be seen that W2.5 and W5 have a higher IFT than W1. IFT can be reduced by using surfactants. For this study, a 0.1$\%$ of sodium dodecyl benzene sulfonate (Sigma-Aldrich) was dissolved in a 5 Mol NaCl brine to lower the IFT to 39.8 mN/m. A lower IFT is expected to increase fluid mobility and decrease the water saturation of the chip, thereby reducing the final precipitation amount. Similarly, NaCl was dissolved in a 50$\%$ + propan$-2-$ol 50$\%$ mixture. Since NaCl solubility is negligible in propan$-2-$ ol, 2.8 mol is close to the solubility limit. That mixture has a low IFT of 27.5 mN/m. Due to its higher vapor pressure compared to water (6.05 kPa to 3.17 kPa), this mixture is expected to have not only higher mobility but also to evaporate faster. Such compositional modifications provide insights into strategies for mitigating salt precipitation. Ammonia-water solutions can facilitate precipitation kinetics for the storage of ammonia with hydrogen. Three different concentrations of NaCl are considered to cover the precipitation kinetics of the NaCl fraction. Due to the low solubility of NaCl in ammonia solution, 3.5 mol is close to the saturation limit. Ammonia solutions (A1, A2.5, and A3.5) have a lower IFT than water + NaCl solutions. They also have a higher vapor pressure (37.3 kPa). Therefore, crystallization is expected to occur more quickly. For comparison, a Stargard brine (S2.4 fluid) has been taken from the Stargard Szczeciński GT-2 geothermal plant located in northwest Poland.
	
	\begin{table}[h]
		\centering
		\caption{Composition of tested experimental solutions, including NaCl concentration of brines, interfacial tension (IFT) for hydrogen at 5 MPa, and flow rate availability for each fluid.}
		\label{tab:IFT_Flow}
		\footnotesize	
		\begin{tabular}{l l l r r c c c c}
			\toprule
			Composition & NaCl  (Mol)  & ID & IFT (mN/m)   & 200 ml/min & 400 ml/min & 1300 ml/min \\
			\midrule
			Water & 1      & W1    & 72.8   &    \checkmark   & \checkmark   & \checkmark \\
			Water &2.5     & W2.5  & 83.5   & \checkmark & \checkmark & \checkmark \\
			Water &5      & W5    & 83.4   &    \checkmark & \checkmark & \checkmark \\	
			Water + SDBS$^*$ 0.1$\%$ &  5         & SDBS5  & 39.8   &    \checkmark & \checkmark & \checkmark \\				
			Water 50$\%$ + propan$-2-$ol 50$\%$ & 2.8      & I2.8  & 27.5   &    \checkmark & \checkmark & \checkmark \\
Water 75$\%$ + Ammonia 25$\%$ & 1      & A1    & 60.9   &   			\checkmark & \checkmark & \checkmark \\
			Water 75$\%$ + Ammonia 25$\%$ & 2.5      & A2.5    & 64.4   &     \checkmark & \checkmark & \checkmark \\	
			Water 75$\%$ + Ammonia 25$\%$ & 3.5      & A3.5    & 67.3   &    \checkmark & \checkmark & \checkmark \\		
			Stargard Brine & 2.4  & S2.4  & 79.5   &    \checkmark & \checkmark & \checkmark \\
			Water 75$\%$ + Ammonia 25$\%$$^\dagger$ & 4      & A4    & 48.7   &      &  &  \\	 
			\bottomrule
			$^*$ sodium dodecyl benzene sulfonate\\
            $^\dagger$ CO$_2$ at 5 MPa and 300 ml/min flow-rate
		\end{tabular}
	\end{table}

\subsection{Crystallization imaging and digital image processing}	
\begin{figure}
	\includegraphics[width=0.95\textwidth]{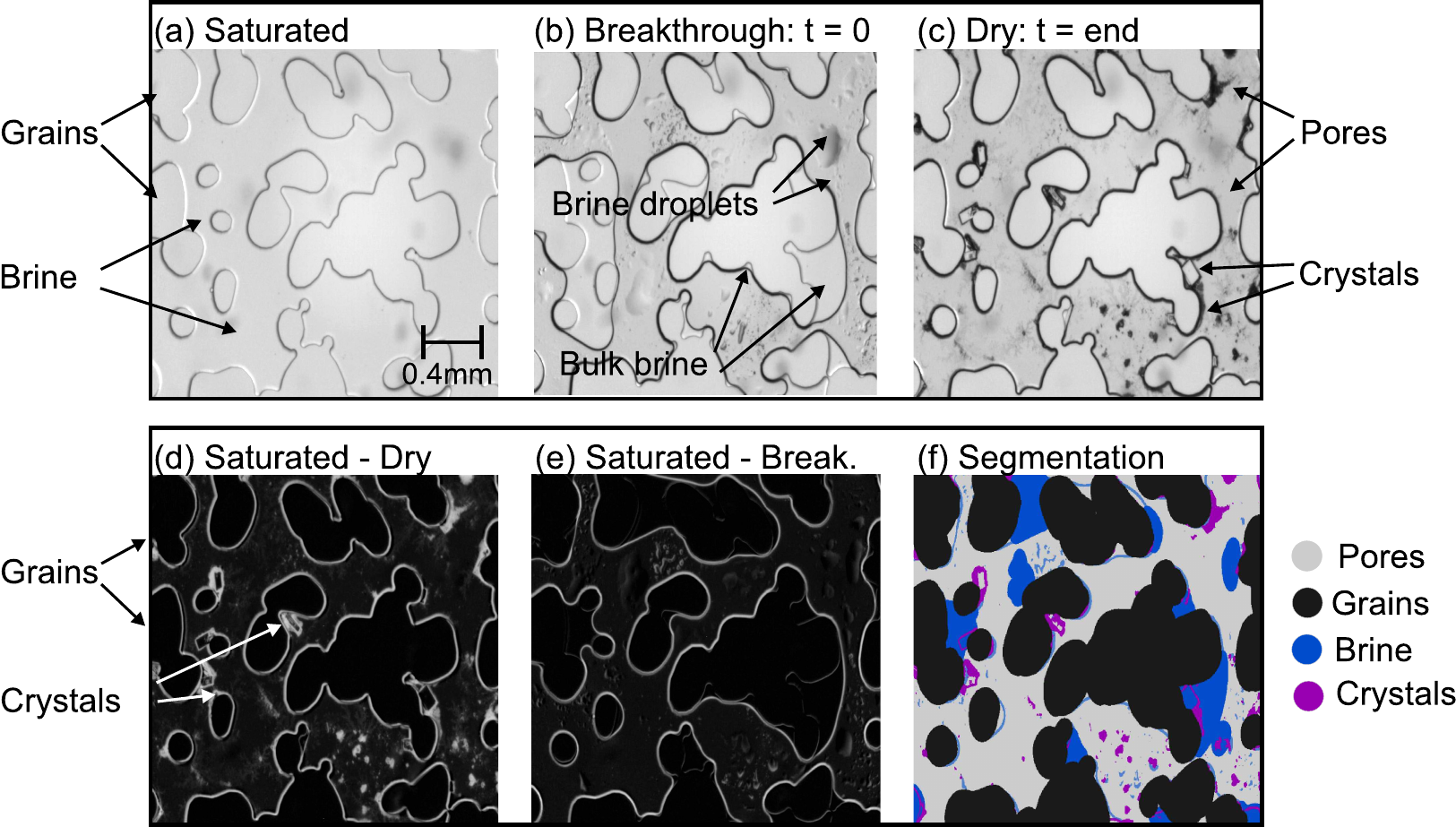}
	\caption{High-resolution microscopic images of a microfluidic chip: (a) fully saturated with brine, (b) partially desaturated after H$_2$ breakthrough, (c) completely dried after brine dryout, with individual phases indicated by arrows (W5 fluid, 400 ml/min flow rate, 10 mm from chip inlet); (d) differential image highlighting crystal positions; (e) differential image highlighting brine positions; (f) segmented image with detected phases from (b) and (c) marked in distinct colors.}
	\label{fig:Segmentation}
\end{figure}

Figure~\ref{fig:Segmentation} (a--c) presents microscopic images illustrating the key stages of the crystallization process at a single location. The images are subtracted from the background and normalized. Figure~\ref{fig:Segmentation} (a) depicts a fully saturated chip at the start of the experiments. Due to the similar refractive indices of water (1.33) and borosilicate glass (1.47), the intensity and contrast differences between glass grains and brine are subtle and indistinct. When the hydrogen pressure drops along the chip, a gas breakthrough occurs. This rapid process displaces part of the brine from the chip. The flow rate is then set to the desired value. The typical stabilization time is on the order of seconds. In this study, we define the breakthrough as the onset of the evaporation process. Fig.~\ref{fig:Segmentation} (b) shows a microscopic image after breakthrough. Residual brine forms bulk brine pools attached to glass grains and droplets within the pore network. The hydrogen angle of refraction is $n\approx 1$, a significant contrast and intensity change between the gas phase and the brine or glass is visible. Hydrogen flow induces the evaporation of residual brine saturation within the porous microfluidic chip. The induction time, defined as the interval between the start of the process ($t=0$) and the detection of the first crystallite (as a proxy for stable, developed nuclei), is one of the key parameters describing precipitation and growth kinetics. At a certain point, the chip is fully dried, concluding the process, and the opaque crystals, visible as dark spots in Figure~\ref{fig:Segmentation}, are clearly distinguishable from grains and pores. The total crystallization time, defined as the duration from the detection of initial crystallites to complete drying, is the second parameter characterizing crystallization kinetics.

In this study, saturation, breakthrough, and images after drying are recorded at 10 adjacent positions along the chip. Each image covers an area of 2$\times$2 mm.  In addition, time scans where the evaporation process is observed are recorded (process between $t=0$ and $t=end$) at a single position, 10 mm from the chip inlet (at the chip's middle point). 

The primary focus of the analysis is to quantify crystal growth and brine evaporation using a custom MATLAB algorithm for image segmentation. Fig.~\ref{fig:Segmentation} (d-f) shows the main steps of the segmentation procedure. First, the image of a fully saturated chip is subtracted from the fully dry chip (Fig.~\ref{fig:Segmentation} (d)). Grains appear as dark patches with near-zero intensity (black in the RGB map) and are outlined by bright white boundaries, distinguishing them from the pore network due to their distinct intensity and color. In contrast, crystals manifest as intense white spots. By segmenting images based on color and intensity, we generate masks for grains and crystals. Similarly, subtracting a fully saturated image from a post-breakthrough image (Figure~\ref{fig:Segmentation} (e)) reveals bulk brine as low-intensity black patches, while water droplets appear as intense white spots. Masks for both bulk brine and water droplets are then created. Figure~\ref{fig:Segmentation} (f) displays the masks generated for each phase of the microfluidic chip. These masks enable the calculation of the network’s porosity, defined as $\phi = A_p / A$, where $A$ is the total chip area and $A_p$ is the pore area. Similarly, brine saturation and crystal fraction are calculated as $S_w = A_w / A_p$ and $X_c = A_c / A_p$, where $A_w$ and $A_c$ represent the areas of brine and crystals, respectively.

\begin{figure}
	\includegraphics[width=0.95\textwidth]{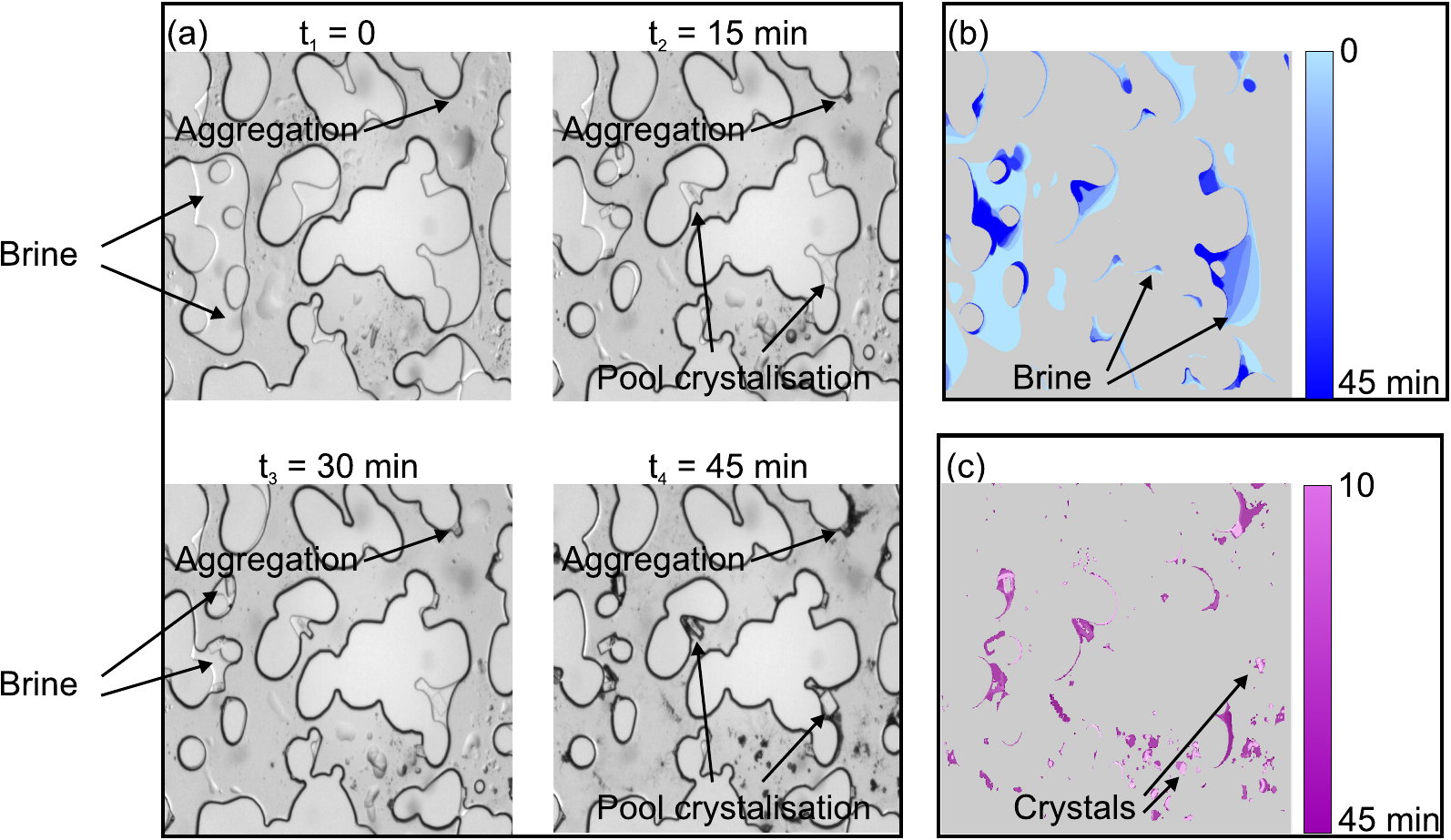}
	\caption{Time evolution of brine evaporation and crystal growth for W5 fluid at a 400 ml/min flow rate: (a) time-sequence images showing the onset of evaporation, crystallization, and final dryout; (b) stacked segmented images illustrating brine evaporation, with the colorbar indicating the time from evaporation onset to final dryout; (c) stacked segmented images depicting crystal growth, with the colorbar representing the time from detection of initial crystallites to final dryout.}
	\label{fig:TimeDryoutandCrystalGrowth}
\end{figure} 

To analyze crystal growth kinetics, we generate a crystal mask from the final crystal distribution and subtract the time-scan image from the fully saturated image. Intensity and color analysis are conducted within the initial crystal mask, tracing backward from $t = t_{\text{end}}$ to $t = 0$, resulting in progressively smaller crystal masks. Similarly, for brine segmentation, we perform the analysis within an initial brine mask (at $t = 0$) until $t = t_{\text{end}}$. Notably, image segmentation requires images spanning the entire process. Figure~\ref{fig:TimeDryoutandCrystalGrowth} (a) displays four time-sequence images of the evaporation process for W5 fluid at a 400 mL/min hydrogen flow rate, highlighting two distinct crystal growth mechanisms (indicated by arrows): (1) growth within an initial brine pool, forming rectangular single crystals, and (2) aggregation into a polycrystalline porous structure outside the brine pool, fed by capillary flow from a neighboring pool.

In the current microfluidic chip design, brine access is limited to residual brine, unlike in a reservoir, where sustained capillary flow can lead to extensive crystallization, posing a limitation for studying crystallization processes. Figure~\ref{fig:TimeDryoutandCrystalGrowth} (b--c) illustrates the time evolution of segmented brine and crystal masks, with color maps indicating the duration from crystallization onset to complete dryout for brine and from the detection of initial crystallites to complete evaporation for crystals. Brine predominantly accumulates on the right side of grains, as hydrogen flow tends to displace brine from the left, though the system cannot track hydrogen flow direction or local velocity across grain surfaces or within the porous network’s constrictions. Additionally, brine evaporates more slowly on the concave surfaces of the porous structure.

\section{Results}

\subsection{Spatial distribution and morphology of brine saturation} 
\begin{figure}[h!]
    \centering
    \includegraphics[width=0.9\textwidth]{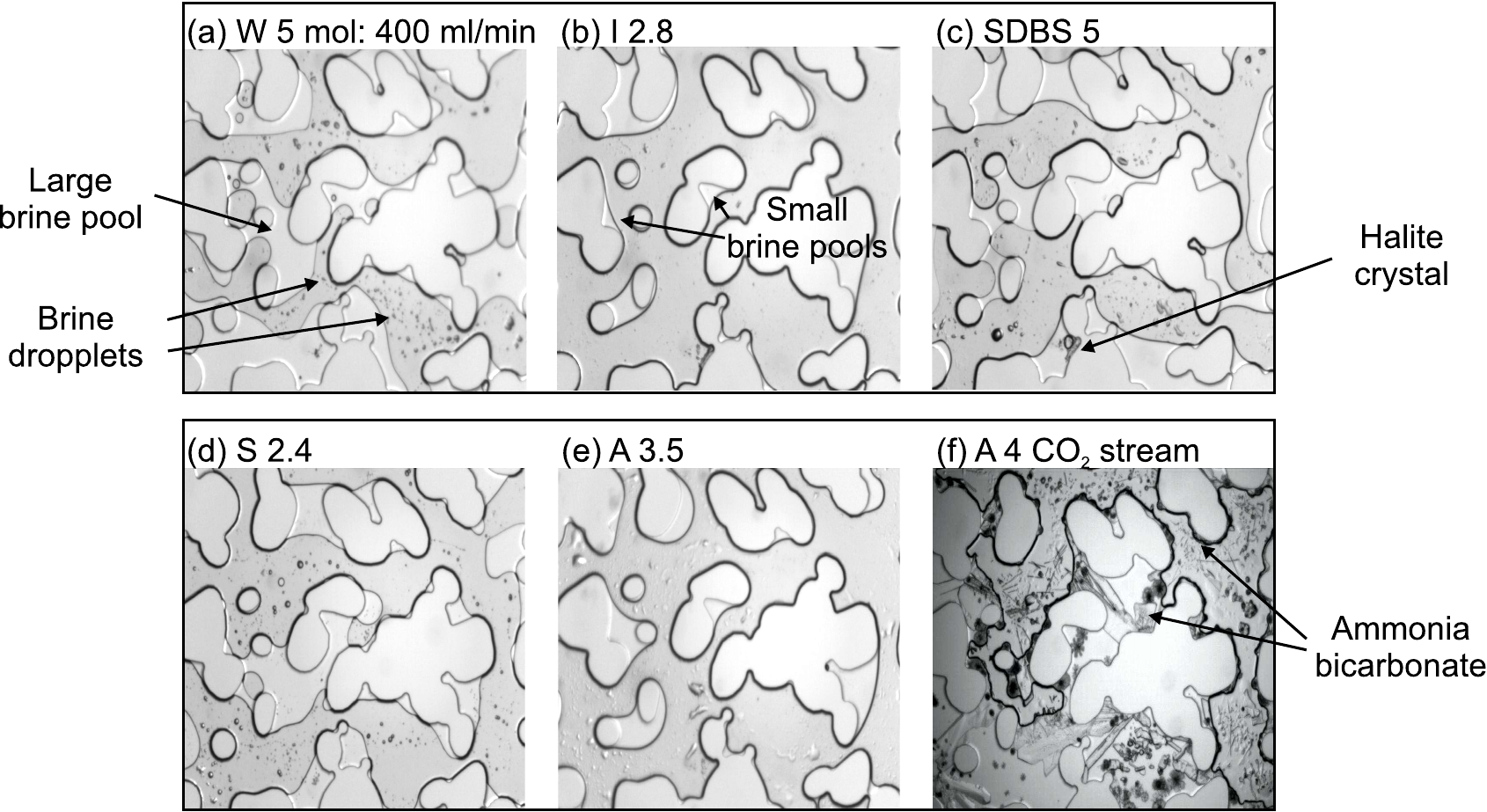}
    \caption{Initial brine saturation after H$_2$ breakthrough ($t = 0$, 400 mL/min flow) for fluids listed in Table~\ref{tab:IFT_Flow}: (a--e) microscopic images of various fluids, with arrows indicating brine pools and droplets (a--b), rapid crystallization (c), and large brine pools (d--e); (f) comparative image of ammonia solution (4 mol NaCl, A4 fluid) after CO$_2$ breakthrough, with arrows marking massive ammonia bicarbonate crystallization.}
    \label{fig:SwInitialVariousSamples}
\end{figure}

The spatial distribution of residual brine significantly influences crystallization kinetics, with higher initial saturation correlating with increased crystal fraction. The evaporation time and subsequent crystallization depend on both the brine distribution and the geometry of the porous network. Figure~\ref{fig:SwInitialVariousSamples} (a--e) illustrates the initial brine distribution for various fluids (listed in Table~\ref{tab:IFT_Flow}) at a hydrogen (H$_2$) flow rate of 400 mL/min, with fluid IDs labeled. For comparison with the carbon dioxide (CO$_2$) injection condition, Figure~\ref{fig:SwInitialVariousSamples} (f) shows the initial brine saturation for a 75\% water + 25\% ammonia solution (4 mol NaCl, fluid A4) under CO$_2$ flow. Notably, W5 fluid (Figure~\ref{fig:SwInitialVariousSamples} (a)) forms large, interconnected brine pools spanning multiple grains, exhibiting high connectivity that supports crystal growth at adjacent grains, with numerous droplets visible in the porous network. In contrast, I2.8 fluid (Figure~\ref{fig:SwInitialVariousSamples} (b)) forms smaller pools adjacent to single grains with fewer droplets, attributable to its lower interfacial tension (IFT), which enhances brine displacement due to higher mobility. Despite its lower IFT, SDBS5 fluid (Figure~\ref{fig:SwInitialVariousSamples} (c)) shows brine saturation comparable to W5, with rapid crystal growth occurring post-breakthrough. The Stargard reservoir brine (S2.4 fluid, Figure~\ref{fig:SwInitialVariousSamples} (d)) exhibits a saturation pattern akin to W5, forming large pools across multiple grains. For the ammonia solution (A3.5 fluid, Figure~\ref{fig:SwInitialVariousSamples} (e)), brine pools are smaller, adjacent to one or two grains, but droplets are larger than in other fluids.

Exposure of ammonia solutions to CO$_2$ (Figure~\ref{fig:SwInitialVariousSamples} (f)) induces massive ammonia bicarbonate crystallization via the reaction \cite{sutter2017solubility}:
\begin{equation}
\text{NH}_3 + \text{HCO}_3^- \rightleftharpoons \text{NH}_2\text{COO}^-,
\end{equation}
forming large crystals that increase brine saturation compared to H$_2$ flow cases. This demonstrates that gas composition, in addition to brine properties, controls the saturation profile and crystallization dynamics. In reservoir rocks, impurities may react with brine or the rock, causing the precipitation of insoluble compounds that affect saturation, crystallization, and ultimately porosity and permeability. In the microfluidic chip, rapid CO$_2$ saturation of brine achieves thermodynamic equilibrium with the gas phase across the chip, promoting uniform crystallization in brine pools.

Brine saturation is influenced by fluid properties, gas composition, and the porous network's structure, leading to local variations across the microfluidic chip. To quantify average saturation across diverse grain distributions, we analyzed 10 positions spanning the chip, providing a qualitative description of the breakthrough process.

\begin{figure}
	\centering	
	\includegraphics[width=1\textwidth]{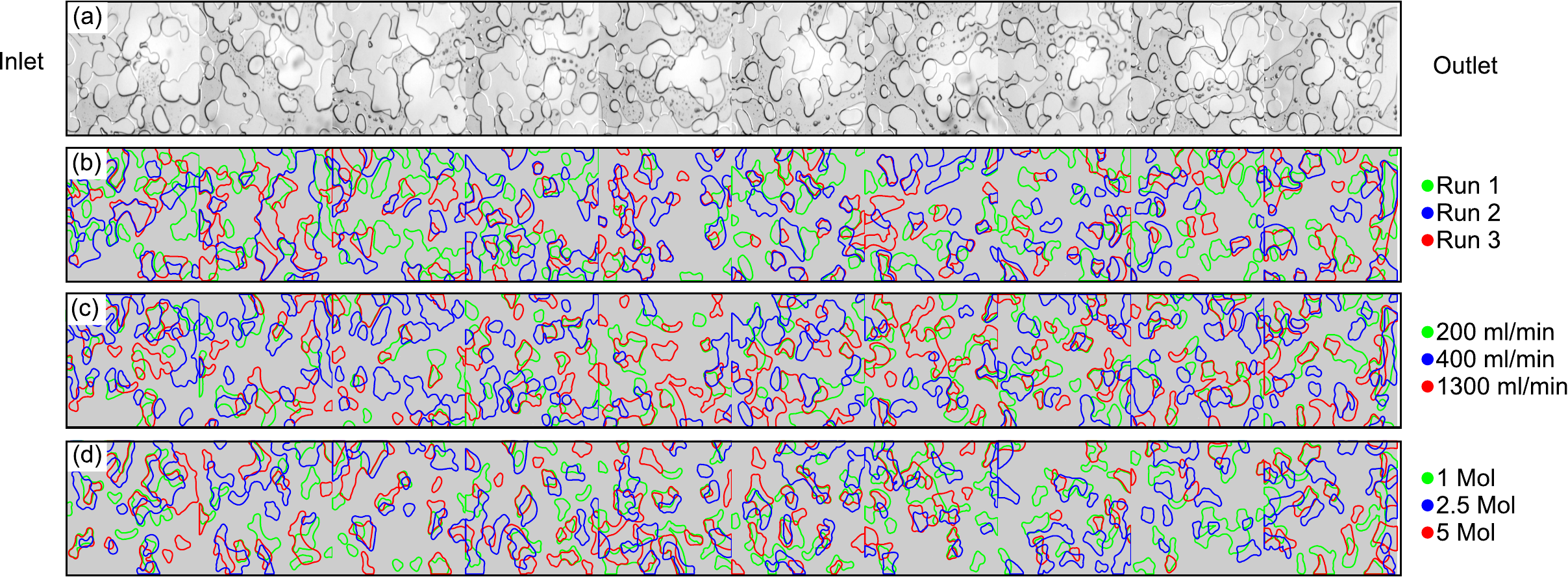}
    \centering
		\caption{Panoramic images of initial brine saturation (20$\times$2 mm, stack of 10 positions) for W5 fluid: (a) raw microscopic image at 400 mL/min flow rate; (b) contour plots of brine pools for three experiments at 400 mL/min, with colors indicating different runs; (c) contour plots of brine pools for three H$_2$ flow rates; (d) contour plots of brine pools for W1, W2.5, and W5 fluids at varying NaCl concentrations and 400 mL/min.}
	\label{fig:PanoramicSwW5Fluid}
\end{figure}

Figure~\ref{fig:PanoramicSwW5Fluid} (a) presents a panoramic image (20$\times$2 mm) of W5 fluid at a 400 mL/min H$_2$ flow rate, constructed from 2$\times$2 mm frames. Slight misalignment in the chip’s transverse orientation causes visible shifts between adjacent frames. The collecting channel appears at the chip outlet, while the distribution channel at the inlet is obscured by tubing connections. Figures~\ref{fig:PanoramicSwW5Fluid} (b--d) display contour plots of bulk brine masks: (b) for three evaporation experiments with W5 fluid at 400 mL/min, revealing significant variability in brine pool size, shape, and position across runs, with no consistent spatial pattern, indicating random hydrogen breakthrough; (c) for different flow rates, showing similar randomness in pool positions and no evident saturation change with flow rate; and (d) for W1, W2.5, and W5 fluids at 400 mL/min, where differences in interfacial tension (IFT) do not significantly alter saturation patterns. The quantitative analysis of brine saturation is presented in the subsequent section.

\begin{figure}
	\includegraphics[width=0.95\textwidth]{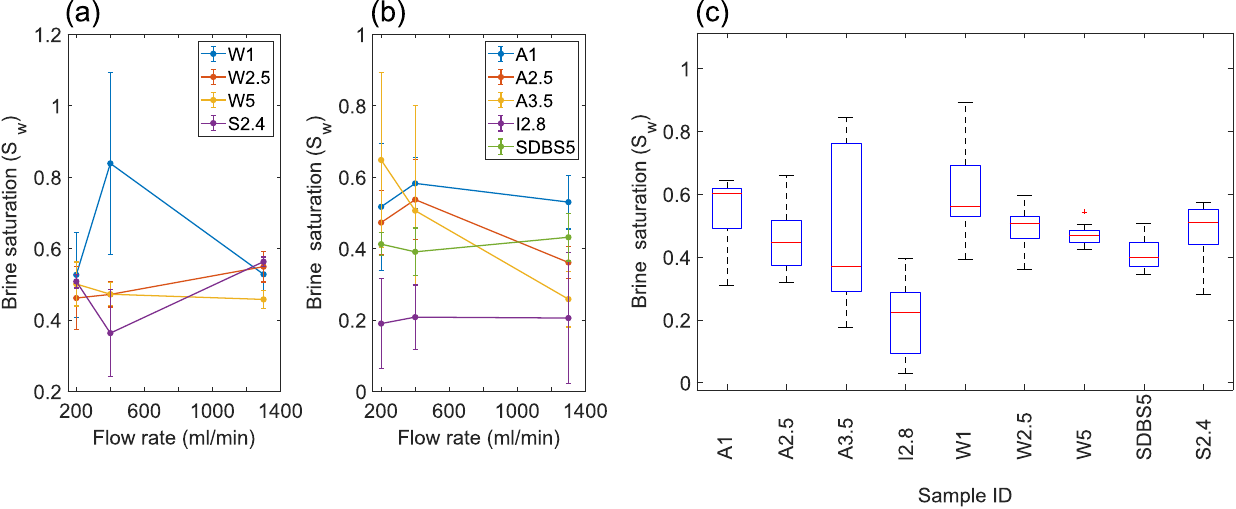}
    \centering
	\caption{Initial brine saturation ($S_w$): (a--b) $S_w$ versus H$_2$ flow rate for two fluid groups (fluid IDs in Table~\ref{tab:IFT_Flow}), with colors distinguishing fluids and points indicating averages with standard deviation whiskers; (c) $S_w$ for each fluid across all flow rates and positions, with red marks for medians, boxes for 25th--75th percentiles, whiskers for non-outlier extremes, and `+' for outliers.}
	\label{fig:SwStatistic}
\end{figure}

We calculated brine saturation ($S_w$) for each position, sample, and flow rate. Figure~\ref{fig:SwStatistic} (a--b) presents $S_w$ as a function of flow rate for two fluid groups, with points indicating average saturation and whiskers representing standard deviation. High variability in saturation is evident, particularly at lower flow rates, where standard deviations occasionally yield unrealistic $S_w$ values exceeding 1. For the first group ( Figure~\ref{fig:SwStatistic} (a)), W2.5 and W5 fluids exhibit comparable average saturation across all flow rates, while W1 and S2.4 show similar saturation at 400 and 1300 mL/min but diverge at lower flows, with W1 displaying significantly higher saturation and S2.4 lower. In the second group (Figure~\ref{fig:SwStatistic} (b)), ammonia solutions A2.5 and A3.5 show reduced $S_w$ at higher flow rates, whereas A1, I2.8, and SDBS5 maintain stable saturation. Notably, I2.8’s lower saturation reflects its high mobility, making it more susceptible to displacement by flowing gas. Similarly, SDBS5 exhibits lower saturation at reduced flow rates compared to other fluids, indicating that flow rate influences $S_w$ selectively, which is modulated by IFT, although random variations often dominate.

Figure~\ref{fig:SwStatistic} (c) illustrates $S_w$ across all flow rates and positions for each fluid, with boxes showing the 25th--75th percentiles, red marks for medians, whiskers for non-outlier extremes, and `+' for outliers. Ammonia solution A3.5 has the lowest average saturation and highest variability, while A1 shows the highest saturation with minimal variation. I2.8 exhibits significantly lower $S_w$ than other fluids, consistent with its high mobility. For W1, W2.5, and W5, higher salinity slightly reduces average saturation. SDBS5’s saturation is lower than W5’s, while S2.4’s is comparable to W2.5’s. IFT significantly impacts saturation: SDBS reduces IFT by approximately 50\%, lowering $S_w$ by 20\%, while propan-2-ol in I2.8 reduces IFT by 70\%, decreasing $S_w$ by 50\%. Comparing W1 and A1, A1 has 10\% higher $S_w$ despite lower IFT, whereas A2.5 and A3.5 show 20\% lower saturation than W2.5 and W5, highlighting the interplay between IFT and fluid composition in controlling saturation.  
	
\subsection{Evaporation and crystal growth dynamics} 		
Following the hydrogen breakthrough, residual brine evaporation begins in the porous microchip. Individual patches evaporate at varying rates due to differences in local H$_2$ advection velocity and brine pool morphology, leading to local fluctuations in saturation. Crystallization initiates when a pool's concentration reaches the solubility limit, exhibiting spatial variations that cause uneven progression of crystal growth.

\begin{figure}[h!]
	\includegraphics[width=0.9\textwidth]{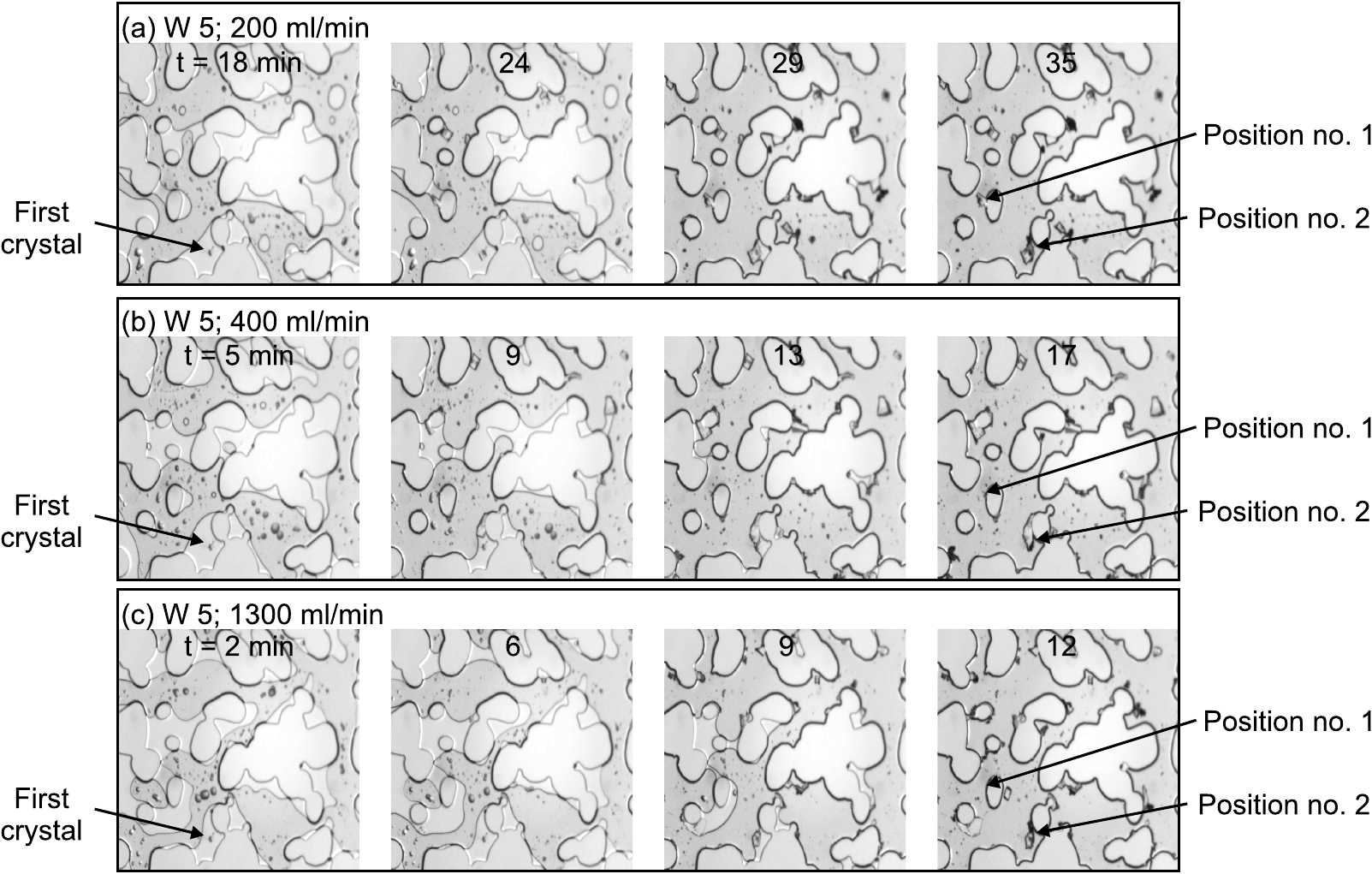}
    \centering
	\caption{Microscopic images of the crystallization process for W5 fluid at varying H$_2$ flow rates of 200, 400 and 1300 ml/min (a--c): left panels show initial crystal crystallite formation, right panels depict the end of evaporation, with upper labels indicating flow rate and central labels specifying the snapshot time from evaporation onset. Arrows on the right highlight the first crystal’s position and points of interest.}
    \label{fig:TimeEvolutionW5FlowRates}
\end{figure}

Figure~\ref{fig:TimeEvolutionW5FlowRates} (a--c) illustrates the dry-out process for W5 fluid across four time steps at different flow rates, with the left side showing initial crystal precipitation and the right side depicting the end of evaporation. Upper labels indicate the time elapsed since the H$_2$ breakthrough. For each flow rate,  crystallite formation consistently begins at approximately the same location (marked by an arrow). Higher flow rates accelerate crystallization and evaporation, reducing the time from crystallite formation to dry-out from 17 minutes at 200 mL/min to 10 minutes at 1300 mL/min. At a specific position (Position 2, marked by an arrow on the right-hand side), a crystal consistently forms by the process's end, though its size and shape vary due to differences in initial brine pool size. In contrast, at Position 1, crystals appear on the left side of the grain at 200 mL/min, are absent at 400 mL/min, and form on the right side at 1300 mL/min, indicating that certain locations favor crystallization while others exhibit more random patterns.
	
\begin{figure}[h!]
	\includegraphics[width=0.9\textwidth]{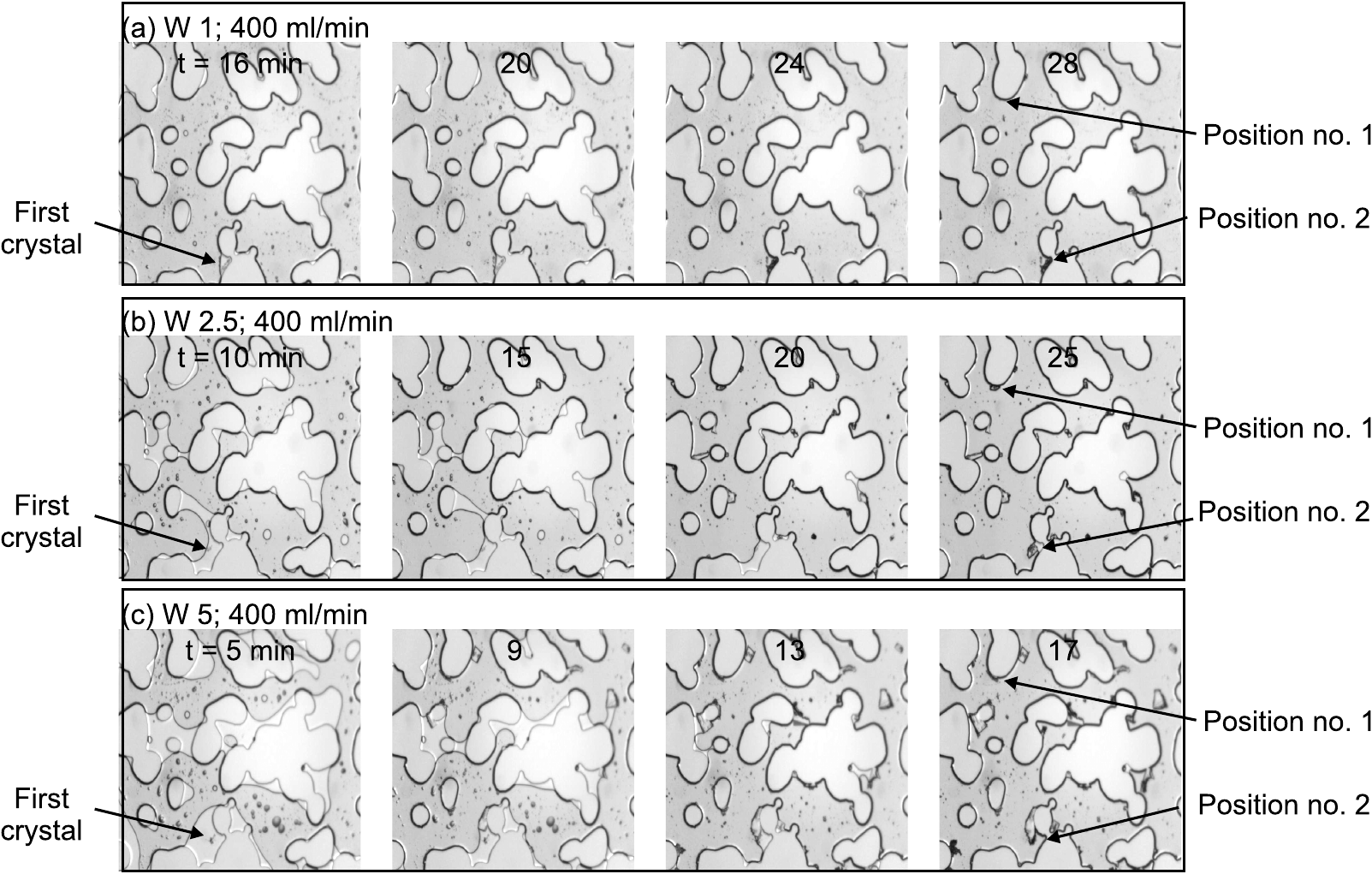}
    \centering
	\caption{Microscopic images of the crystallization process for W1, W2.5, and W5 fluids at a 400 mL/min H$_2$ flow rate (a--c): left panels show initial crystallite formation, right panels depict the end of evaporation, with upper labels indicating the flow rate and central labels specifying the snapshot time from evaporation onset. Arrows on the right highlight the first crystal’s position and points of interest.}
	\label{fig:TimeEvolutionFlow400Concentration}
\end{figure}

Figure~\ref{fig:TimeEvolutionFlow400Concentration} (a--c) illustrates the crystallization process for W1, W2.5, and W5 fluids at a 400 mL/min flow rate, with left panels showing initial crystal crystallite formation and right panels depicting the end of evaporation. Consistent with Figure~\ref{fig:TimeEvolutionW5FlowRates},  crystallite formation initiates at the exact location (marked by an arrow on the left). Higher brine concentrations accelerate the onset of crystallization, as less evaporation is required to reach the solubility limit, and expedite the completion of evaporation. Hydrophilic crystals form a capillary brine film on their surfaces, increasing the evaporation area and thus the evaporation rate. However, the duration from crystallite formation to complete drying is comparable for W1 and W5 fluids (12 minutes) but longer for W2.5 (15 minutes). Lower brine concentrations result in a reduced final crystal fraction, characterized by fewer and smaller crystallites exhibiting random distributions. For example, at Position 1, no crystal forms for W1, whereas crystals are present for W2.5 and W5. At Position 2,  crystallite formation occurs consistently across all fluids, though crystal size and shape vary (Figure~\ref{fig:TimeEvolutionFlow400Concentration}).

The evaporation and crystallization processes are quantitatively characterized at each stage using brine saturation ($S_w$) and crystal fraction ($X_c$) profiles over time. Figure~\ref{fig:TimeProfiles} (a--b) presents these profiles for experiments depicted in Figures~\ref{fig:TimeEvolutionW5FlowRates} and \ref{fig:TimeEvolutionFlow400Concentration}, with solid lines representing $S_w$ and dashed lines indicating $X_c$. In Figure~\ref{fig:TimeProfiles} (a), colored lines correspond to flow rates of 200, 400, and 1300 mL/min, showing similar initial brine saturation (left axis) across all flow rates. Higher flow rates accelerate the onset and completion of crystallization, with the final crystal fraction (right axis) highest at 1300 mL/min and lowest at 400 mL/min, reflecting random local variations in $X_c$ that do not directly correlate with initial $S_w$, as crystallite formation may occur outside the microscope’s field of view. In Figure~\ref{fig:TimeProfiles} (b), profiles for NaCl concentrations of 1, 2.5, and 5 mol/kg show that higher concentrations initiate crystal growth earlier and yield a greater final crystal fraction compared to lower concentrations. The brine saturation profile can occasionally increase or exhibit sudden drops. During flow, brine droplets may detach from brine pools and move downstream, or attach to neighboring pools, altering the measured saturation.

\begin{figure}
	\includegraphics[width=0.9\textwidth]{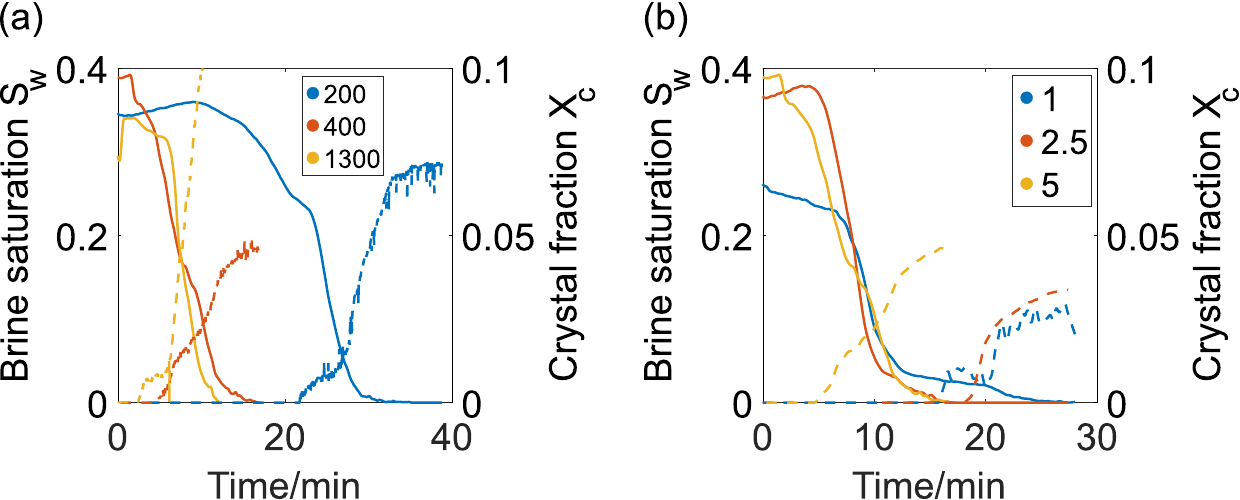}
    \centering
	\caption{Temporal profiles of brine saturation ($S_w$) and crystal fraction ($X_c$) for the crystallization and evaporation processes: (a) profiles for W5 fluid at H$_2$ flow rates of 200, 400, and 1300 mL/min, corresponding to images in Figure~\ref{fig:TimeEvolutionW5FlowRates}, with solid lines representing $S_w$ (left axis) and dashed lines indicating $X_c$ (right axis), where colors denote flow rates and higher flow rates show earlier crystallite formation and higher final $X_c$; (b) profiles for W1, W2.5, and W5 fluids at a 400 mL/min flow rate, corresponding to images in Figure~\ref{fig:TimeEvolutionFlow400Concentration}, with solid and dashed lines for $S_w$ and $X_c$, respectively, where colors indicate NaCl concentrations (1, 2.5, 5 mol/kg) and higher concentrations yield earlier crystal growth and greater $X_c$.}
    \label{fig:TimeProfiles}
	\label{fig:TimeProfiles}
\end{figure} 	

\begin{figure}[h!]
	\includegraphics[width=0.9\textwidth]{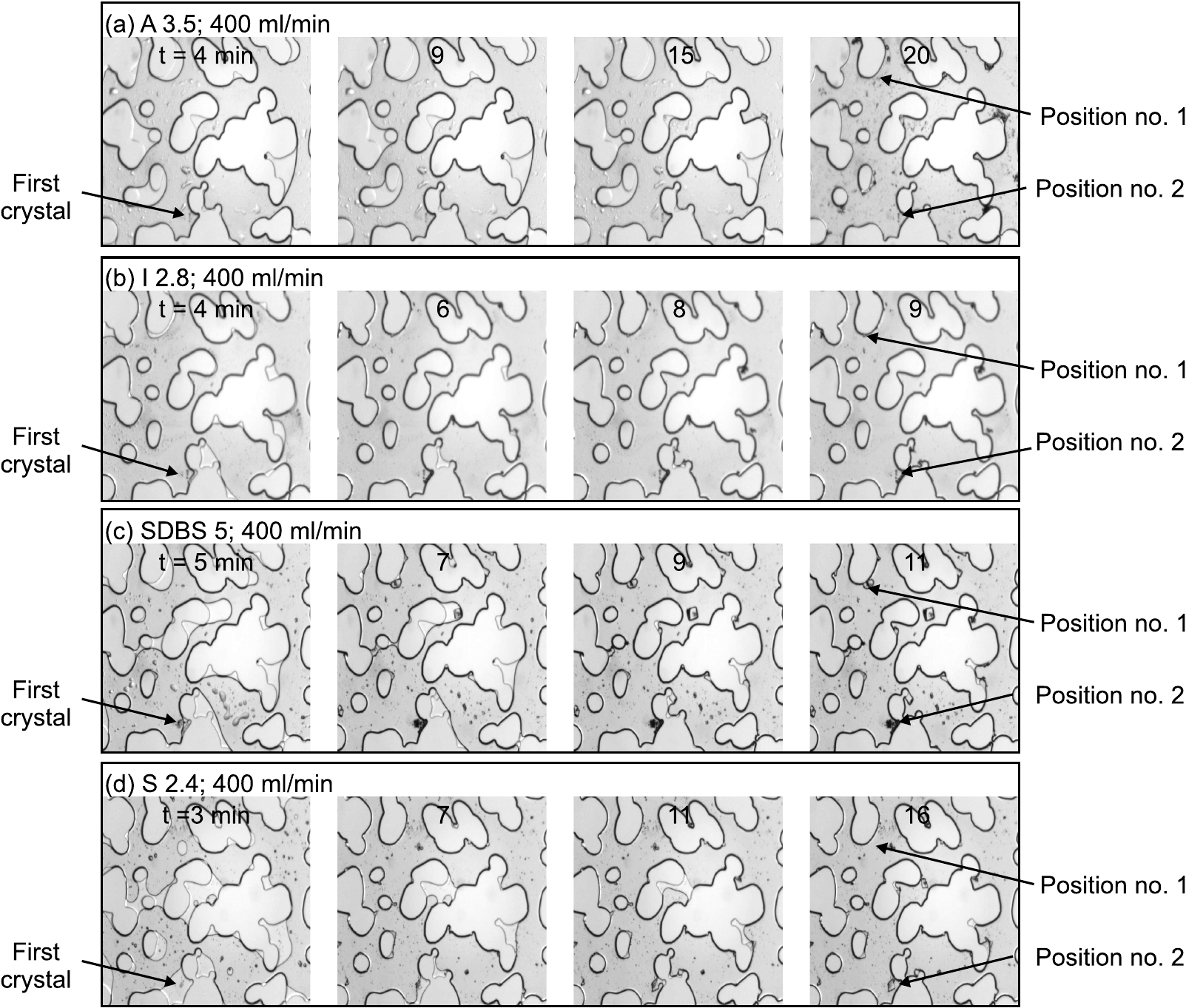}
    \centering
	\caption{Microfluidic images of the crystallization process for A3.5, I2.8, SDBS5, and S2.4 fluids at a 400 mL/min H$_2$ flow rate (a--d): left panels show initial crystallite formation, right panels depict the end of evaporation, with upper labels indicating the flow rate and central labels specifying the snapshot time from evaporation onset. Arrows on the right highlight the first crystal’s position (Position 2) and points of interest (Position 1), where crystal morphology varies from small crystals (A3.5) to large single crystals (I2.8, S2.4) or polycrystalline aggregates (SDBS5).}
	\label{fig:TimeEvolutionSamplesSupplement}
\end{figure}

Crystallization kinetics are analyzed for A3.5, I2.8, SDBS5, and S2.4 fluids at a 400 mL/min hydrogen flow rate, as shown in Figure~\ref{fig:TimeEvolutionSamplesSupplement} (a--d). For A3.5 and I2.8,  crystallite formation begins at 4 minutes, earlier than for W5 (5 minutes), while S2.4 nucleates faster at 3 minutes, and SDBS5 matches W5 at 5 minutes. Crystallization for A3.5 concludes at 20 minutes, with growth persisting for 16 minutes, longer than for W5 (12 minutes) and W2.5 (15 minutes) but ending later than W2.5 and faster than W5. For I2.8, low brine saturation ($S_w$) and a high evaporation rate result in rapid dry-out (9 minutes) and a short crystal growth phase (5 minutes), which is significantly faster than for W5 or W2.5, indicating that alcohol addition reduces crystallization time. Similarly, SDBS5 exhibits a brief 6-minute growth phase and completes drying in 11 minutes. For S2.4, the 13-minute growth phase is shorter than W2.5 but starts earlier.

Crystallite formation consistently initiates at Position 2 for all fluids, though crystal morphology varies: A3.5 produces a small crystal, I2.8 and S2.4 form large single crystals, and SDBS5 yields polycrystalline aggregation. This suggests that initial crystallite formation at a fixed position does not guarantee large crystal sizes, and the type of crystallization (single crystal or aggregation) is independent of salinity, interfacial tension (IFT), or fluid type. At Position 1,  crystallite formation is more random: A3.5 and S2.4 show no crystals, I2.8 forms a small single crystal, and SDBS5 produces a large single crystal, highlighting the stochastic nature of the crystallite formation as a proxy for the nucleation process.

\begin{figure}[h!]
	\includegraphics[width=0.7\textwidth]{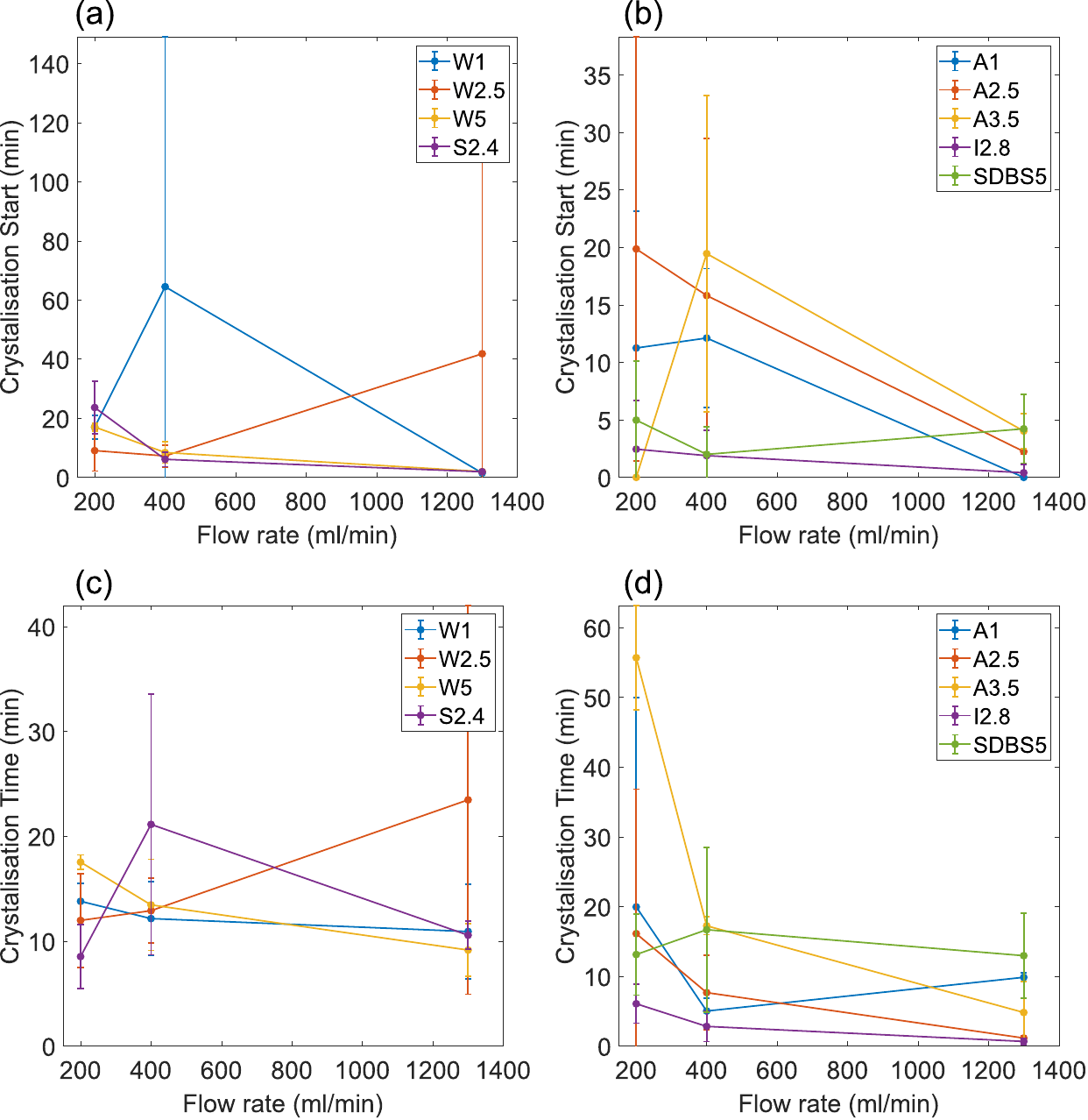}
    \centering
	\caption{Salt precipitation initiation and duration for various experimental fluids. (a--b) First crystallization time for two fluid groups (see Table~\ref{tab:IFT_Flow}), with colors indicating individual fluids (W1, W2.5, W5, S2.4, etc.) and points showing averages with uncertainty whiskers, where higher flow rates reduce crystallite formation time and uncertainty, notably high for W1 at 400 mL/min and W2.5 at 1300 mL/min. (c--d) Crystal growth duration for the same fluid groups, with colors denoting fluids, where higher flow rates generally shorten growth time for W1 and W5 but extend it for W2.5, with significant uncertainty for S2.4 at 400 mL/min and reduced uncertainty at higher flow rates for the second group.}
	\label{fig:TimeStats}
\end{figure}

We analyzed the first crystallite formation time and the duration of the crystal growth stage for various fluids and flow rates, as detailed in Table~\ref{tab:IFT_Flow}. Figure~\ref{fig:TimeStats} (a--b) presents the first crystallite formation time for two fluid groups, while (c--d) shows the crystal growth duration. For clarity, fluids are divided into two groups. In Figure~\ref{fig:TimeStats} (a), W1 at 400 mL/min and W2.5 at 1300 mL/min exhibit high uncertainty in crystallite formation time, whereas other cases show lower uncertainty. Higher flow rates generally reduce the uncertainty in crystallite formation time and accelerate the onset of crystallite formation. Similarly, for the second fluid group (Figure~\ref{fig:TimeStats} (b)), nucleation time shows greater uncertainty at lower flow rates, with earlier growth and reduced uncertainty at higher flow rates.

The crystal growth stage duration, shown in Figure~\ref{fig:TimeStats} (c--d), varies with flow rate. For the first group (Figure~\ref{fig:TimeStats} (c)), higher flow rates shorten growth time for W1 and W5 fluids, while S2.4 shows an extended growth time and high uncertainty at 400 mL/min. In contrast, W2.5 exhibits increased growth time and uncertainty with higher flow rates. For the second fluid group (Figure~\ref{fig:TimeStats} (d)), both growth time and its uncertainty decrease with increasing flow rate, with a notable reduction between 200 and 400 mL/min. These trends indicate that higher flow rates enhance evaporation, thereby accelerating crystal growth.    

\subsection{Spatial distribution and morphology of salt crystals} 	
	
\begin{figure}
	\centering	
	\includegraphics[width=\textwidth]{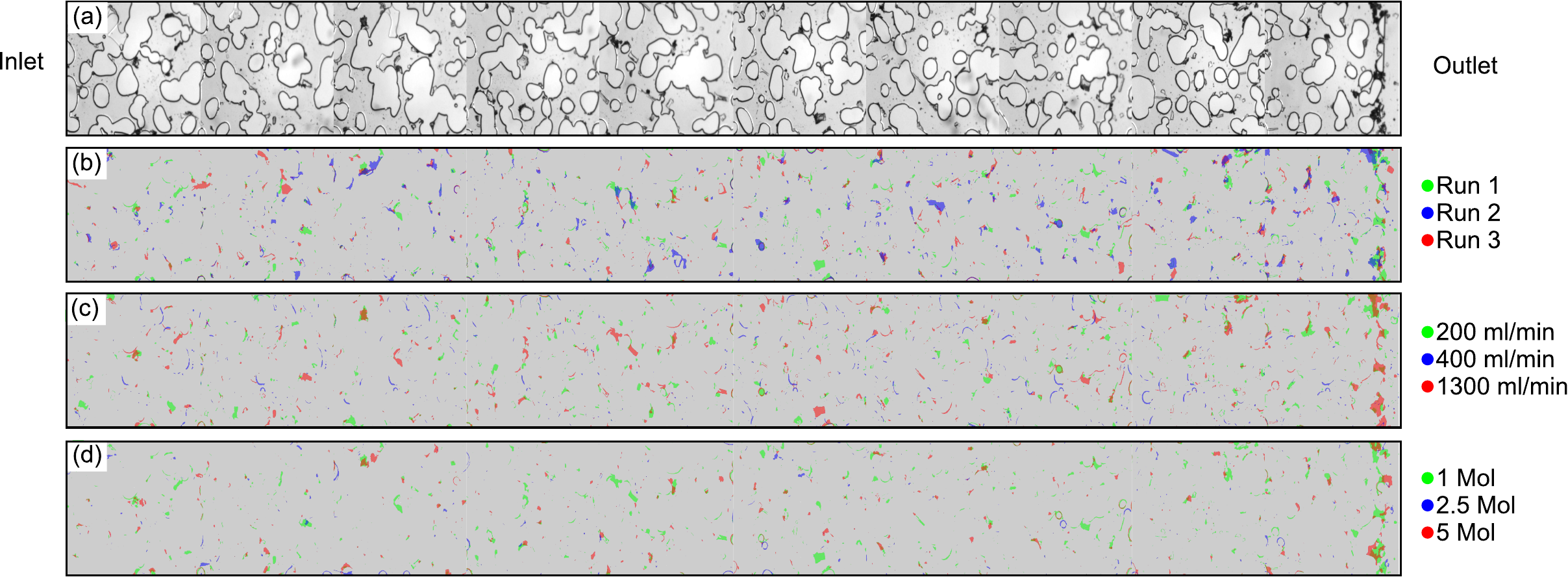}
	\caption{Spatial distribution and morphology of H$_2$-induced salt crystals post-evaporation. Panoramic images (20$\times$2 mm, stack of 10 positions) of final crystal fraction ($X_c$) corresponding to brine saturation in Figure~\ref{fig:PanoramicSwW5Fluid}: (a) raw microscopic image of W5 fluid (5 mol/kg NaCl) at 400 mL/min H$_2$ flow, showing halite as single crystals and aggregates, with extensive precipitation near the chip outlet; (b) crystal masks for three experiments with W5 fluid at 400 mL/min, with colors indicating different runs, highlighting high crystal fraction near the outlet and random distributions with spatial variations; (c) crystal positions for W5 fluid at 200, 400, and 1300 mL/min, with colors denoting flow rates, showing larger crystals at 1300 mL/min near the outlet; (d) crystal positions for W1, W2.5, and W5 fluids (1, 2.5, 5 mol/kg NaCl) at 400 mL/min, with colors indicating concentrations.}
	\label{fig:PanoramicXcW5Fluid}
\end{figure}

The spatial distribution and morphology of salt crystals following brine evaporation are analyzed using methods analogous to those applied for initial brine saturation, offering insights into the interplay among precipitation, evaporation, and the porous network structure. Figure~\ref{fig:PanoramicXcW5Fluid} (a--d) illustrates the crystal fraction ($X_c$) corresponding to the brine distributions in Figure~\ref{fig:PanoramicSwW5Fluid}. Figure~\ref{fig:PanoramicXcW5Fluid} (a) presents a panoramic image (stack of 10 microscopic frames, 20$\times$2 mm) of W5 fluid (5 mol/kg NaCl) after a 400 mL/min hydrogen flow, revealing halite precipitation as single crystals and aggregates. Large crystals predominantly adhere to grain surfaces, while crystals from brine droplets form within porous channels. Extensive precipitation occurs near the chip outlet’s collection channel, where a large cavity retains significant brine post-breakthrough, promoting higher crystal fractions. Some regions with residual brine, impacted by curvature or crystal clogging, resist evaporation and are excluded from temporal analysis to ensure accurate kinetic measurements.

Figure~\ref{fig:PanoramicXcW5Fluid} (b) displays crystal masks for three distinct experiments with W5 fluid at 400 mL/min, with colors indicating crystallites from consecutive runs. A consistently high crystal fraction is observed near the chip outlet across all runs, reflecting favorable conditions for precipitation in the collection channel. However, crystal distributions exhibit randomness, with spatial variations: Run 3 shows large crystal clusters near the chip inlet, Run 2 forms large crystals in the middle or near the outlet, and Run 1 displays evenly distributed large crystals. Even in shared growth sites, crystal sizes vary, driven by the stochastic nature of nucleation and the random distribution of brine pools post-hydrogen breakthrough, which underscores the probabilistic influence of the porous network on crystal morphology.

Figure~\ref{fig:PanoramicXcW5Fluid} (c--d) further explores the impact of flow rate and concentration. In (c), the final crystal coverage for W5 fluid at 200, 400, and 1300 mL/min exhibits random distributions, with larger crystals forming at 1300 mL/min, particularly near the chip outlet, due to enhanced evaporation rates that accelerate the growth of early-nucleating crystals compared to lower flow rates. In (d), crystal distributions for W1, W2.5, and W5 fluids (1, 2.5, and 5 mol/kg NaCl) at 400 mL/min reveal lower crystal coverage for lower concentrations, with W5 forming larger aggregates than W1 or W2.5. This trend reflects the higher initial brine saturation in W5, which promotes greater crystal growth, highlighting the critical role of concentration in determining final crystal fraction and morphology.
    
\begin{figure}[h!]
	\includegraphics[width=\textwidth]{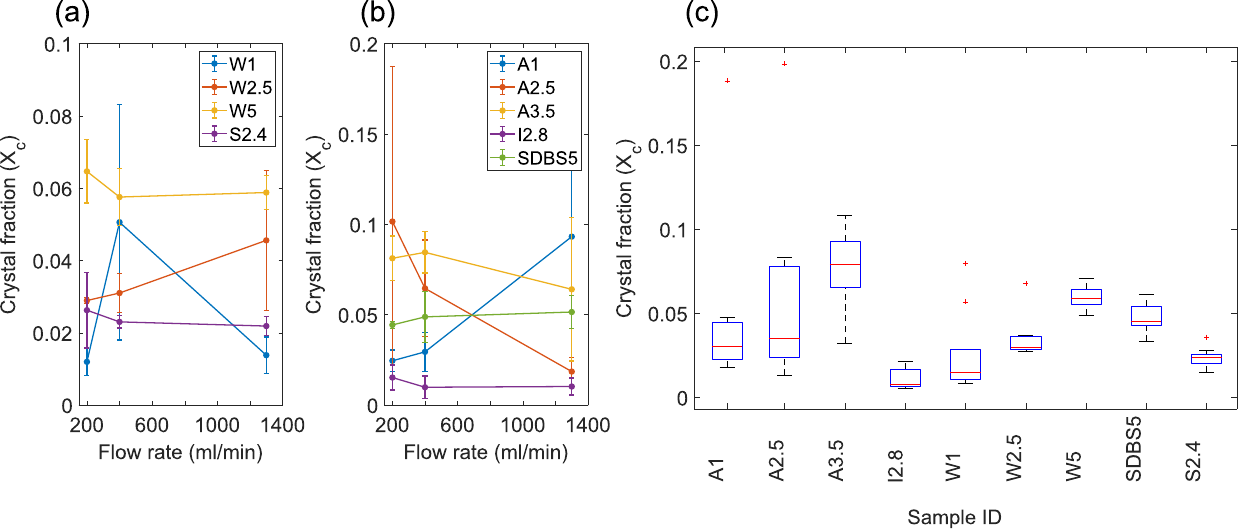}
	\caption{Salt crystal fraction analysis across experimental fluids and H$_2$ flow rates. (a--b) Crystal fraction ($X_c$) versus flow rate (200, 400, 1300 mL/min) for two fluid groups (see Table~\ref{tab:IFT_Flow}), with colors indicating fluids (W1, W2.5, W5, A1, A2.5, A3.5, I2.8, SDBS5, S2.4), points showing average $X_c$ per run, and whiskers representing uncertainty. (c) Crystal fraction ($X_c$) for each fluid across all flow rates and positions, with boxes showing 25th--75th percentiles, red marks for medians, whiskers for non-outlier extremes, and '+' for outliers.}
	\label{fig:XcStatistic}
\end{figure}

The crystal fraction ($X_c$) for various fluids and experimental conditions is quantitatively analyzed to elucidate the effects of H$_2$ flow rate and fluid composition, paralleling the analysis of brine saturation ($S_w$). Figure~\ref{fig:XcStatistic} (a--b) presents $X_c$ as a function of flow rate for two fluid groups (defined in Table~\ref{tab:IFT_Flow}), with points representing the average $X_c$ across positions for each run and whiskers indicating uncertainty. For W1, W2.5, and W5 fluids, higher NaCl concentrations (1, 2.5, 5 mol/kg) correlate with increased $X_c$, reflecting the direct influence of initial brine saturation. Specifically, W5 and S2.4 show a slight decrease in $X_c$ with increasing flow rate, though uncertainty dominates, while W2.5 exhibits a clear increase in $X_c$ with flow rate, with significant uncertainty at 1300 mL/min. W1 displays variable $X_c$ behavior, with high uncertainty at 400 mL/min, consistent with elevated $S_w$ observed in Figure~\ref{fig:SwStatistic} (a), highlighting the linkage between brine saturation and crystal formation.

In Figure~\ref{fig:XcStatistic} (b), the second fluid group reveals distinct trends: I2.8, with alcohol, exhibits the lowest $X_c$ (decreasing with flow rate), attributed to its reduced $S_w$. SDBS5 shows a lower $X_c$ than A3.5 or A2.5 but increases with flow rate, while A1’s $X_c$ is lower than A2.5 and A3.5 at low flow rates but rises significantly at 1300 mL/min. Conversely, A2.5’s $X_c$ decreases with flow rate, with high uncertainty at 200 mL/min, surpassing A3.5’s uncertainty. A3.5’s $X_c$ also decreases with flow rate, indicating that ammonia-based fluids respond variably to flow dynamics, influenced by their interfacial tension (IFT) and evaporation rates.

Figure~\ref{fig:XcStatistic} (c) presents $X_c$ across all flow rates and positions for each fluid, with boxes showing the 25th--75th percentiles, red marks for medians, whiskers for non-outlier extremes, and `+' for outliers. Ammonia-based fluids show distinct behaviors: A3.5 achieves the highest $X_c$ (0.08), exceeding W5 (0.06), with higher uncertainty than A2.5 (0.035) and A1 (0.03). A1’s $X_c$ surpasses W1 (0.015) and is comparable to W2.5 (0.03), indicating that ammonia enhances crystal growth. In contrast, I2.8’s $X_c$ (0.008) is the lowest, reflecting a 70\% reduction in $S_w$ compared to W2.5, which suppresses crystallization and potential chip clogging. SDBS5 reduces $X_c$ by 30\% (0.046) compared to W5, a less pronounced effect than alcohol. S2.4’s $X_c$ (0.024) is similar to W2.5 (0.035), suggesting that alternative salts mimic NaCl’s impact on crystal growth. These results underscore the critical role of the initial $S_w$ in determining $X_c$, modulated by additives like ammonia (enhancing growth), alcohol (suppressing growth), and SDBS (moderately reducing growth).

\section{Discussion}

\subsection{Fluid Composition Controls on Brine Distribution and Crystallization Dynamics}

Our experimental results demonstrate that fluid composition fundamentally governs both brine evaporation and salt precipitation patterns under hydrogen injection in porous media through multiple interconnected mechanisms. In our experiments, the interfacial tension (IFT) directly influences brine pool formation during the initial hydrogen breakthrough phase and subsequently determines the spatial distribution and connectivity of residual brine available for crystallization.

\subsubsection{IFT effects on brine pool morphology}
High-IFT fluids (W-fluids, A-fluids, and Stargards brine) consistently generate large, interconnected brine pools that span multiple grain surfaces. The enhanced connectivity among growth sites fosters a network effect, where crystal growth propagates across adjacent sites through nucleation and growth on secondary substrates \cite{nooraiepour2021probabilistic, masoudi2022effect,nooraiepour2025three,nooraiepour2021probabnuc}. This process results in increased crystal fractions and broader precipitation coverage, contributing to the self-reinforcing feedback loops characterized as salt self-enhancing growth dynamics. The large pool volumes also provide a sustained brine supply for prolonged crystal growth, enabling the formation of well-developed halite aggregates \cite{masoudi2024understanding,dkabrowski2025surface}.

Conversely, low-IFT fluids: I2.8 (\SI{27.5}{\milli\newton\per\meter}) and SDBS5 (\SI{39.8}{\milli\newton\per\meter}) exhibit markedly different behavior. Their enhanced mobility during displacement results in more efficient brine removal, producing smaller, isolated pools with significantly reduced connectivity. This isolation limits crystal growth to individual crystallite formation sites and prevents the formation of large aggregate structures. The rapid displacement and subsequent fast evaporation characteristic of these systems create shorter crystal growth windows, effectively suppressing large-scale crystallization and substantially reducing the risk of pore-network clogging.

The relationship between IFT and final crystal fraction is evident when comparing low-IFT fluids to W-fluids and Stargard brine. For instance,  SDBS5 despite identical NaCl concentrations (5 molar), a \SI{\sim50}{\percent} reduction in IFT via surfactant addition leads to a \SI{\sim20}{\percent} decrease in crystal fraction when compared to W5 fluid. Similarly, the alcohol-containing I2.8 fluid, exhibiting a reduction in IFT \SI{\sim70}{\percent} compared to W2.5 fluid and shows a reduction in final crystal coverage \SI{\sim50}{\percent}, highlighting a robust quantitative correlation between interfacial properties and the extent of precipitation.

\subsubsection{Chemical composition impact beyond IFT}

Ammonia-containing fluids reveal that chemical composition effects extend beyond simple IFT modifications, introducing additional complexity to saturation and crystallization behavior. Ammonia-containing fluids exhibit lower residual saturation compared to equivalent-molarity pure NaCl solutions (W-fluids). Specifically, A2.5 and A3.5 fluids show reduced residual saturation relative to their NaCl counterparts (W2.5 and W5), whereas the A1 fluid exhibits slightly higher residual saturation than W1. Additionally, the variability in residual saturation for high-salt ammonia samples is greater than that observed for high-salinity W-fluids.

The elevated vapor pressure of ammonia solutions (\SI{37.3}{\kilo\pascal} versus \SI{3.17}{\kilo\pascal} for pure water) accelerates evaporation, though this effect interacts with changes in displacement efficiency during breakthrough. Interestingly, despite generally lower residual saturation, ammonia-based fluids consistently reach higher final crystal fractions than equivalent-concentration NaCl solutions. Notably, A3.5 achieves the highest overall crystal coverage (\SI{0.08}{}) among all tested fluids.

From a mass balance perspective, lower crystal coverage might be anticipated. However, this discrepancy likely arises from brine displacement within the chip during evaporation, as brine droplets may detach from the main pool and flow downstream. For ammonia fluids, crystallization occurs more rapidly than for water-based fluids, often commencing immediately after breakthrough, with crystals forming almost instantaneously in some experiments. The shorter process duration and earlier onset of crystallization reduce the volume of displaced brine, thereby increasing the final crystal fraction. Nevertheless, the current system does not permit quantitative measurement of the displaced brine volume.

\subsubsection{Flow rate dependencies}
Varying hydrogen flow rates reveals distinct fluid-dependent behaviors \cite{jangda2024subsurface,johnson2024impact,wang2024influence,johnson2025multicycle} that provide insight into the underlying physical processes controlling crystallization. Higher flow rates universally accelerate the onset of crystallite formation and reduce crystal growth durations by enhancing mass transfer and evaporation rates. However, the magnitude and character of this response vary systematically with fluid properties.

Low-IFT fluids (I2.8, SDBS5) demonstrate the most pronounced flow rate sensitivity, exhibiting rapid dry-out with abbreviated growth phases that intensify at higher flow rates. This behavior reflects their inherently high mobility and limited brine retention, making them particularly responsive to enhancements in advective transport. In contrast, high-IFT fluids (W5, S2.4) sustain longer growth phases even at elevated flow rates, indicating that their larger, more stable brine pools provide sufficient mass transfer resistance to buffer against increases in flow rate.

Ammonia-based fluids, particularly A3.5, exhibit intermediate flow rate sensitivity but maintain extended growth times relative to aqueous NaCl solutions. This behavior, combined with their enhanced final crystal fractions, suggests that ammonia's influence on crystallization kinetics involves both thermodynamic effects (e.g., altered solubility and vapor pressure) and kinetic effects (modified nucleation barriers and surface interactions) that operate across the entire evaporation timeline.

\subsubsection{Probabilistic nucleation and pore geometry controls}

The observed stochastic distribution of crystallite formation sites (as proxies for nucleation events) across the microfluidic network, punctuated by recurring preferred nucleation positions, reveals the complex interplay between fluid chemistry and porous network geometry. While fluid properties determine the overall tendency for crystallization through residual saturation levels and evaporation kinetics, the porous network geometry introduces probabilistic controls that can either amplify or suppress local precipitation \cite{goh2010stochastic,nooraiepour2021probabilistic}.

This pore geometric influence manifests as position-dependent variations in crystal size, morphology, and occurrence probability, even for identical fluid compositions. The persistence of certain preferred crystallite formation sites across different fluids suggests that local geometric features, such as pore throat constrictions, grain surface curvature, and flow field convergence zones, can create favorable conditions for crystal formation that transcend fluid-specific effects. However, the stochastic nature of crystallite formation (nucleation events) suggests that thermodynamic-geochemical conditions remain the primary determinant of overall crystallization behavior, with geometric effects serving as a secondary modulating influence \cite{nooraiepour2021probabnuc}.

\subsection{Comparative Analysis of Hydrogen versus Carbon Dioxide Induced Salt Precipitation}

The distinct characteristics of H\textsubscript{2} and CO\textsubscript{2} injection systems extend beyond their molecular identities, encompassing differences in thermophysical properties, chemical reactivity, and transport mechanisms that govern salt precipitation behavior. A key distinction between the two systems lies in their chemical interactions with saline and ammonia-containing brines. Hydrogen exhibits complete chemical inertness across all tested brine compositions (e.g., NaCl and ammonia-based solutions), leading to precipitation driven solely by physical processes, namely evaporation-induced supersaturation and subsequent crystal growth in concentrated residual brines. In contrast, CO\textsubscript{2} displays significant chemical reactivity, particularly with ammonia-containing brines, initiating sequential precipitation reactions that alter both the quantity and composition of precipitated phases. The primary reaction pathway involves CO\textsubscript{2} dissolution, carbonic acid formation, and acid-base chemistry with ammonia, as described by the following equations \cite{sutter2017solubility}:

\begin{align}
\ce{CO2 + H2O &<=> H2CO3} \label{eq:carbonic} \\
\ce{H2CO3 + NH3 &<=> NH4+ + HCO3-} \label{eq:bicarbonate} \\
\ce{NH4+ + HCO3- &<=> NH4HCO3(s)} \label{eq:precipitation}
\end{align}

These reactions produce ammonium bicarbonate and carbonate phases alongside halite, substantially increasing total crystal coverage and accelerating precipitation kinetics. Our prior microfluidic studies \cite{dkabrowski2025microfluidic} demonstrated that CO\textsubscript{2}-ammonia systems can achieve extensive pore blockage due to bicarbonate precipitation, a phenomenon not observed in H\textsubscript{2} systems. The rapid equilibration of CO\textsubscript{2} with brines promotes uniform crystallization across brine pools, in contrast to the spatially heterogeneous, evaporation-limited precipitation observed in H\textsubscript{2} systems.

The contrasting thermophysical properties of H\textsubscript{2} and CO\textsubscript{2} create different displacement mechanisms and post-breakthrough saturation profiles. Table~\ref{tab:H2_CO2_comparison} summarizes key property differences relevant to brine displacement and evaporation processes.

\begin{table}[h!]
\centering
\caption{Comparative thermophysical properties of H\textsubscript{2} and CO\textsubscript{2} relevant to brine displacement and salt precipitation (conditions: \SI{5}{\mega\pascal}, \SI{25}{\degreeCelsius}).}
\label{tab:H2_CO2_comparison}
\begin{tabular}{lccc}
\toprule
Property & H\textsubscript{2} & CO\textsubscript{2} & Ratio (CO\textsubscript{2}/H\textsubscript{2}) \\
\midrule
Density (\si{\kilogram\per\cubic\meter}) & 4.1\cite{wei2023correlations} & 207.8\cite{Laesecke2017} & 51 \\
Dynamic Viscosity (\si{\micro\pascal\second}) & 9.0\cite{wei2023correlations} & 24.8\cite{Laesecke2017} & 2.8 \\
IFT with brine (\si{\milli\newton\per\meter})$^*$ & 72.8 & 55-65 & 0.8 \\
Diffusion coefficient in water (\si{\square\meter\per\second})$^{**}$ & \num{4.5e-9}\cite{kallikragas2014high} & \num{1.9e-9}\cite{cadogan2014diffusion} & 0.4 \\
Solubility in water (mol/L) & \num{7.8e-4}\cite{zhu2022accurate}& 0.034\cite{enick1990co2} & 44 \\
\bottomrule
\multicolumn{4}{l}{$^*$W1 fluid composition; $^{**}$at \SI{25}{\degreeCelsius}}
\end{tabular}
\end{table}

CO\textsubscript{2}'s significantly higher density (\SI{51}{\times} greater) and viscosity (\SI{2.8}{\times} greater) result in more stable displacement fronts and enhanced sweep efficiency during breakthrough. The higher viscosity provides better displacement control, reducing viscous fingering and creating more uniform saturation profiles. Additionally, CO\textsubscript{2}'s lower interfacial tension with brine (approximately 15--20\% reduction compared to H\textsubscript{2}) enhances its ability to displace brine from smaller pores and create more efficient drainage patterns.

These combined effects result in systematically lower post-breakthrough brine saturations for CO\textsubscript{2} systems, as documented in our previous microfluidic investigations \cite{dkabrowski2025microfluidic, dkabrowski2025surface}. However, this relationship appears paradoxical when considering final precipitation outcomes: despite higher residual saturations, H\textsubscript{2} systems generally produce lower final crystal fractions than CO\textsubscript{2} systems (excluding reactive precipitation scenarios).

The contrasting transport properties of H\textsubscript{2} and CO\textsubscript{2} create different mass transfer characteristics that influence evaporation dynamics and precipitation patterns. CO\textsubscript{2}'s lower IFT with brine enhances displacement efficiency and creates thinner brine films with larger surface-area-to-volume ratios, accelerating evaporation rates and supersaturation development. Additionally, CO\textsubscript{2}'s substantially higher water solubility greater than H\textsubscript{2}) enables more extensive gas-brine equilibration and enhanced moisture transport capacity within the flowing gas phase. These transport advantages explain the more rapid and spatially uniform brine removal observed in CO\textsubscript{2} systems compared to the slower, more heterogeneous evaporation characteristic of H\textsubscript{2} systems.

The temporal evolution of crystallization exhibits notable differences between gas types. In our experiments, CO\textsubscript{2} systems with ammonia-containing brines showed rapid, widespread precipitation following breakthrough, whereas H\textsubscript{2} systems demonstrated more spatially and temporally heterogeneous precipitation patterns. For CO\textsubscript{2}-ammonia systems, the chemical reactions (Equations~\ref{eq:carbonic}--\ref{eq:precipitation}) can trigger precipitation within minutes of gas-brine contact, creating extensive crystal coverage. In contrast, H\textsubscript{2} systems rely solely on evaporation-driven supersaturation, resulting in more variable precipitation/growth timing that depends on local evaporation rates and initial brine distribution.

The spatial distribution of precipitated phases reveals important differences in precipitation patterns between H\textsubscript{2} and CO\textsubscript{2} systems, with implications for pore-scale clogging behavior. H\textsubscript{2}-induced precipitation produces discrete crystal deposits localized to sites of initial brine accumulation, as crystal growth is constrained by the finite dissolved solids available within individual brine pools. This localized nature limits the spatial extent of precipitation and reduces the likelihood of forming interconnected crystal networks. This suggests different risk profiles for reservoir-scale operations, although direct scaling remains challenging. H\textsubscript{2} systems may primarily face localized precipitation effects near injection points where evaporation rates are highest.

\subsection{Engineering Implications and Environmental Considerations for UHS}

The microfluidic observations from this study offer insights into the environmental aspects of engineering underground hydrogen storage (UHS) systems. However, careful consideration of scaling relationships and geological complexity is essential for translating pore-scale findings into reservoir applications.

Our experimental findings reveal that H\textsubscript{2}-induced salt precipitation is governed by mechanisms that share similarities with, yet diverge from, those in CO\textsubscript{2} systems, particularly in their dependence on physical rather than chemical processes. The purely evaporation-driven nature of H\textsubscript{2} precipitation can inform operational strategies. As discussed, IFT modifications may offer a promising avenue for precipitation control. Our experiments demonstrate that alcohol additives (I2.8 fluid) reduce crystal fraction by approximately \SI{50}{\percent} compared to equivalent-salinity aqueous solutions, while surfactant addition (SDBS5) achieves \SI{30}{\percent} reduction. These findings suggest that deliberate formation water modification could provide effective near-wellbore protection against precipitation-induced permeability damage. However, implementing such strategies requires careful consideration of environmental regulations, long-term chemical stability, and potential interactions with reservoir rocks \cite{belhaj2020effect,hou2024long}. Alcohol-based additives may pose environmental concerns and could alter H\textsubscript{2} purity during withdrawal, while surfactants require evaluation of adsorption behavior and thermal stability under reservoir conditions.

Flow rate optimization represents another viable mitigation strategy. Higher injection rates consistently reduce crystallite formation times and crystal growth durations across all tested fluids, suggesting that maintaining adequate flow velocities could prevent extensive precipitation buildup.

The transition from two-dimensional microfluidic observations to three-dimensional reservoir behavior involves several critical scaling considerations that limit direct extrapolation of our findings. Pore network connectivity in reservoir rocks is fundamentally three-dimensional, providing multiple flow pathways that can accommodate precipitation without complete blockage \cite{guo2025microfluidic,roman2025microfluidics}. Our microfluidic chips constrain flow to essentially two-dimensional channels, potentially overestimating clogging risk by eliminating bypass opportunities available in real rock systems. Three-dimensional systems introduce additional complexity through enhanced capillary and gravity-driven mechanisms, which amplify the self-enhancing nature of salt precipitation \cite{nooraiepour2025three}. The stochastic crystallite formation patterns observed in our experiments suggest that reservoir heterogeneity can either amplify or suppress precipitation effects, depending on the specific combination of pore structure, flow distribution, and chemical conditions. Thermal effects, largely absent in our ambient-temperature experiments, become significant at reservoir depths. Elevated temperatures will accelerate both evaporation and crystal growth kinetics while altering phase behavior and solubility relationships. These thermal effects require systematic investigation under reservoir-relevant conditions.

\section{Conclusions}

This study provides a systematic pore-scale visualization of salt precipitation during hydrogen injection, establishing fundamental mechanistic differences from extensively studied CO$_2$ systems that have critical implications for underground hydrogen storage design. Our microfluidic investigation of 81 high-pressure experiments reveals that hydrogen-induced precipitation operates through physical processes—evaporation and capillary trapping—producing discrete, localized deposits with generally lower clogging risk than reactive CO$_2$ systems. In contrast, CO$_2$ injection into ammonia-containing brines triggers chemical precipitation of ammonium bicarbonate, accelerating crystallization and creating interconnected crystal networks that severely compromise permeability.

Three key findings emerge with direct operational relevance. First, interfacial tension controls both residual brine distribution and precipitation severity: low-IFT chemical additives (alcohols, surfactants) reduce crystal coverage by up to 50\% through enhanced brine mobility, offering a practical mitigation strategy for near-wellbore protection. Second, hydrogen's high mobility creates breakthrough patterns with heterogeneous brine distributions, complicating flow predictability but potentially enabling operational control through flow-rate optimization. Third, ammonia-containing brines paradoxically increase crystal fractions despite lower interfacial tension, demonstrating that chemical composition effects extend beyond simple interfacial property modifications—a finding critical for co-storage scenarios where hydrogen displaces stored ammonia in salt caverns.

These mechanistic insights establish that gas-specific risk assessments are essential for underground hydrogen storage—CO$_2$-analog approaches are insufficient and potentially misleading. The high mobility and chemical inertness of hydrogen create precipitation patterns fundamentally distinct from CO$_2$ systems, requiring tailored monitoring, prediction, and mitigation strategies. Site-specific chemical assessments must account for storage history, including residual additives in depleted reservoirs and potential ammonia interactions in salt caverns, as these profoundly alter precipitation pathways.

The demonstrated effectiveness of chemical additives in suppressing precipitation opens promising pathways for engineering solutions. Strategic deployment of low-IFT fluids during injection, combined with flow-rate optimization, could minimize near-wellbore damage while maintaining injectivity. Implementing these strategies necessitates a systematic evaluation via three-dimensional porous media, reservoir thermodynamic conditions, environmental compatibility, long-term chemical stability, and potential impacts on hydrogen purity during withdrawal cycles.

\section*{Data Availability}
The data supporting the findings of this study are openly accessible in the RODBUK repository at \href{https://doi.org/10.58032/AGH/BDK0YN}{https://doi.org/10.58032/AGH/BDK0YN}.

\section*{Acknowledgments}						
This research was supported by the project ``Solid and Salt Precipitation Kinetics During CO$_2$ Injection into Reservoirs,'' funded by the Norway Grants (Norwegian Financial Mechanism 2014--2021) under grant number UMO-2019/34/H/ST10/00564, and partially funded by the EEA and Norway Grants under grant number NOR/POLNORCCS/AGaStor/0008/2019-00. MN and MM acknowledge support from the Norwegian Centennial Chair (NOCC) for the project ``Understanding Coupled Mineral Dissolution and Precipitation in Reactive Subsurface Environments.''

\subsection*{CRediT Authorship Contribution} 
KD: Conceptualization, Methodology, Software, Investigation, Validation, Data Curation, Visualization, Writing - Original Draft, Writing - Review \& Editing.  MN: Conceptualization, Formal analysis, Validation, Writing - Original Draft, Writing - Review \& Editing. MM: Formal analysis, Writing - Review \& Editing.

\subsection*{Conflicts of Interest}
The authors declare that they have no conflicts of interest that could influence the publication of this article.
    

\printbibliography
\end{document}